\documentclass[%
 reprint,
 amsmath,amssymb,
 aps,
floatfix
]{revtex4-2}

\usepackage{graphicx}
\usepackage{dcolumn}
\usepackage{bm}

\usepackage{hyperref}
\usepackage{cleveref}

\begin{document}


\title{Effect of hyperons on \\
$f$-mode oscillations in Neutron Stars}

\author{Bikram Keshari Pradhan} 
\author{Debarati Chatterjee}%
 \email{debarati@iucaa.in}
\affiliation{Inter-University Centre for Astronomy and Astrophysics, \\
Pune University Campus,\\
Pune - 411007, India
}%

\date{\today}

\begin{abstract}
As the densities in the interior of neutron stars exceed those of terrestrial nuclear experiments, they provide a scope for studying the nature of dense matter under extreme conditions. The composition of the inner core of neutron stars is highly uncertain, and it is speculated that exotic forms of matter such as hyperons may appear there. Gravitational waves emitted by unstable oscillation modes in neutron stars contain information about their interior composition and therefore allow us to probe the interior directly. Recently, a systematic investigation of $f$-mode oscillations in neutron stars revealed the role of the uncertainty in nuclear saturation parameters, particularly the effective nucleon mass, on its frequency. In this work, we study the influence of the appearance of hyperons on $f$-mode oscillation frequencies and therefore on the emission of gravitational waves. We also speculate whether a future detection of $f$-mode frequencies could provide a possibility of probing the presence of hyperons in the neutron star core. 
\end{abstract}

\keywords{neutron stars, gravitational waves, f-modes, dense matter, hyperons}                             
                              
\maketitle


\section{Introduction}  
\label{sec:intro} 
Neutron stars (NS) are the stellar remnants left behind when a massive star (with mass $8M_{\odot}$ to 25 $M_{\odot}$) undergoes a supernova explosion. The density of a neutron star core can reach up to 2-10 times the nuclear saturation density and provides an excellent astrophysical laboratory to study the physics of cold ultra-dense matter, beyond the reach of present terrestrial experiments. Though the theory of nuclear interactions close to saturation density is well understood, one must resort to neutron stars  to explore the physics of cold ultra-dense matter.
\\

Neutron stars are observable throughout the electromagnetic spectrum, and many macroscopic properties like mass and radius can be derived from the multi-wavelength electromagnetic observations. The maximum mass can be precisely determined from relativistic effects for neutron stars in binary (\cite{Demorest2010,Antoniadis2013,Cromartie,Fonsecate}), but the radius cannot be precisely determined from X-ray data. Future observations from the recently launched NICER   (Neutron star Interior Composition Explorer) mission  \cite{NICER} and the future eXTP  (enhanced x-ray Timing and Polarimetry Mission) are expected to improve the measurements of radii to higher accuracy  \cite{eXTP}. 
\\

The variation of pressure with density or Equation of State (EoS) of dense matter is a key entity in modelling its global properties. Given an EoS, global observables like the mass and radius of a NS can be obtained by solving hydrostatic equilibrium equations. Comparison with astrophysical data can then help us to constrain EoS models. At the ultrahigh densities in the NS core, the Fermi energy of the constituent nucleons exceeds the in-medium mass of heavier baryons and favours their appearance. As the time scale associated with weak processes is much smaller than the time scale associated with a neutron star, violation of strangeness conservation due to weak interaction leads to the appearance of strange baryons, known as hyperons.
\\

The appearance of new degrees of freedom (such as hyperons) in the system are expected to be accompanied by a lowering of the pressure and a softening of the EoS. The EoS controls the maximum mass and other global properties: it is well known that a softer equation of state corresponds to a lower maximum mass. There are many studies that were carried out with different EoS models including hyperons, and in most of the cases the maximum mass obtained was within the range 1.3-1.8 $M_{\odot}$ (see  \cite{Ramos,Schulze,GM1,Schaffner96}). However, recent astrophysical observations indicate large NS masses $\sim 2 M_{\odot}$. It has been consequently understood, through pioneering works by \cite{ChatterjeeNPA,ChatterjeePRC} followed by many others \cite{Vidana}, that models with repulsive hyperon-hyperon interaction can produce a stiffer equation of state at high densities and can explain the large mass observations. 
\\

One may describe the nuclear EoS using the phenomenological Relativistic Mean Field (RMF) model, in which the baryons interact through the exchange of mesons. In this model, the nuclear coupling constants are obtained by fitting them to saturation properties of nuclear matter. However, extrapolation of saturation nuclear data to high densities and isospin asymmetric matter in neutron stars leads to uncertainties.  The effect of the  uncertainties in the nuclear parameters on NS observables
for a purely nucleonic system has been studied recently \cite{Hornick}. Hyperon-nucleon interaction couplings are determined by symmetry relations of the quark model or fitted to available potential depths from hypernuclear experiments. Hyperon-hyperon coupling constants can similarly be obtained by fitting them to hyperon-hyperon potential depths from double hypernuclei data (\cite{Vidana}, \cite{Schaffner93}). The role of the uncertainty in hyperon-nucleon potential depths and the conditions for appearance of hyperons in massive neutron stars has already been systematically investigated \cite{ChatterjeeNPA}. It was shown that one may obtain a maximum mass of 2$M_{\odot}$ by producing a stiffer EoS, through repulsive hyperon-hyperon vector interactions \cite{ChatterjeePRC} or including quark phase at higher densities  \cite{Klahn}.
\\

NSs emit gravitational waves (GW) in both isolated and binary systems. Any non-axisymmetric perturbation in NSs can result in emission of GWs. These GWs contain the signature of the internal composition of the NS.  In a binary merger, NSs exert strong tidal forces on one another, and the (dimensionless) tidal deformability $\bar{\Lambda}$ depends upon their compactness  \cite{AbbottPRL119}. This can help us to constrain the NSs radii and hence the EoS  \cite{AbbottPRL121}. Analysis of tidal deformability from the event GW170817 was able to set limits on the radius of a $M = 1.4M_{\odot}$ NS in the range of $R_{1.4M_{\odot}}=$ 12-13.5 km and its tidal deformability to $\bar{\Lambda}_{1.4M_{\odot}} <$ 720 \cite{AbbottPRX}. Detection of GWs from binary NS merger,  GW170817, and electromagnetic counterpart GRB 170817A  \cite{AbbottAJL848} marks the beginning of a new era in multi-messenger astronomy opening up a new window to the GW Universe.
\\

An oscillating neutron star (isolated or a merger remnant) can have several quasi-normal modes, classified according to the restoring force that brings  the system back to equilibrium (\cite{Cowling,Schmidt,Thorne}), e.g., fundamental ($f$) modes, pressure ($p$) modes, gravity ($g$) modes, rotational  ($r$) modes or pure space-time ($w$) modes. The frequency associated with different modes depends upon the interior composition of the star. Thus the observation of mode frequencies can be used to probe the core composition of a neutron star. Among the different oscillation modes, $f$-modes are very interesting, as they are predicted to produce significant amounts of gravitational radiation through the Chandrasekhar-Friedman-Schutz (CFS) mechanism when NS is unstable \cite{Glampedakis} and the frequencies lie within the sensitivity band of the upcoming generation of GW detectors.
\\

In this work, we investigate how the appearance of hyperons affects the stellar $f$-mode frequencies for a non-rotating star in the framework of the RMF model. We extend a recent systematic investigation of the role of nuclear saturation parameters on the mode oscillations for a nucleonic system \cite{Sukrit} to include hyperonic degrees of freedom. This paper is organized in the following way. In \Cref{sec:formalism}, we will discuss the model used to calculate EoS, how the model parameters are obtained, and determine the global properties from  EoS. We will also discuss how the $f$-mode frequencies are obtained. In \Cref{sec:results} and \Cref{sec:asteroseismology}  we present our results and in Sec. V we summarize our conclusions.

\section{Formalism}
\label{sec:formalism}

\subsection{Microscopic Description: RMF Model}
For this investigation, we adopt the RMF theoretical model introduced in \Cref{sec:intro} to describe the beta-equilibrated and charge-neutral hadronic matter. The NS core is assumed to be composed of baryons (all baryons in baryon octet ) and leptons ($e^-$ and $\mu^-$). In this model, baryon-baryon interaction is mediated by the exchange of scalar   ($\sigma$), vector   ($\omega$) and isovector ($\rho$) mesons and the hyperon-hyperon interaction is mediated by two  additional mesons, strange scalar   ($\sigma^*$) and strange vector ($\phi$) mesons \cite{Schaffner96}. We also include the vector-isovector ( $\omega-\rho$ ) interaction via the $\Lambda_{\omega}$ coupling \cite{Hornick}. The Lagrangian density for the hadronic phase is given by:
\\

\begin{eqnarray}
     \mathcal{L} &=&\sum_B  \bar{\psi}_{_B}  (i\gamma^{\mu}\partial_{\mu}-m_{_B}+g_{\sigma_ B}\sigma-g_{\omega_ B}\gamma_{\mu}\omega^{\mu}-g_{\rho_B}\gamma_{\mu} \vec{I_B}.\vec{\rho}^{\mu})\psi_{_B} \nonumber \\
     &+&\frac{1}{2}  (\partial_{\mu} \sigma \partial^{\mu}\sigma - m_{\sigma}^2 {\sigma}^2) 
     -U_{\sigma}+\frac{1}{2}m_{\omega}^2 \omega_{\mu}\omega^{\mu}-\frac{1}{4} \omega_{\mu \nu}\omega^{\mu \nu} \nonumber \\
     &-&\frac{1}{4}  (\vec{\rho}_{ \mu \nu}.\vec{\rho}^{\mu \nu}-2 m_{\rho}^2 \vec{\rho}_{\mu}.\vec{\rho}^{\mu})+\Lambda_{\omega} (g_{\rho_N}^2  \vec{\rho}_{\mu} .\vec{\rho}^{\mu}) \  (g_{\omega_N}^2 \omega_{\mu}\omega^{\mu})  \nonumber \\ 
     &+&\mathcal{L}_{YY}+\mathcal{L}_{\ell} 
     \label{eqn:lagr}
\end{eqnarray}
where,
 \begin{eqnarray}
     U_{\sigma}&=&\frac{1}{3}b m_N   (g_{\sigma_N} \sigma)^3+\frac{1}{4}c   (g_{\sigma_N} \sigma)^4 \nonumber\\
     \mathcal{L}_{YY}&=&\sum_Y  \bar{\psi}_{_Y}  (g_{\sigma^*_Y} \sigma^*-g_{\phi_ Y}\gamma_{\mu}\phi^{\mu})\psi_{_Y}+\frac{1}{2}m_{\phi}^2 \phi_{\mu}\phi^{\mu}\nonumber\\
     && \ \ -\frac{1}{4} \phi_{\mu \nu}\phi^{\mu \nu}  +\frac{1}{2}  (\partial_{\mu} \sigma^* \partial^{\mu}\sigma^* - m_{\sigma^*}^2 {\sigma^*}^2) \nonumber\\
    \mathcal{L}_{\ell}&=&\sum_{\ell=\{e^- , \ \mu^-\}}  \bar{\psi}_{\ell}  (i\gamma^{\mu}\partial_{\mu}-m_{\ell}){\psi}_{\ell} \nonumber
\end{eqnarray}

Within the mean-field approximation, we replace meson fields by their mean value, which is obtained assuming baryons are in the ground state. The expectation values of the meson fields are now represented by `$\sigma , \omega_0 ,\rho_{03} ,\sigma^*,\phi_0$'. From the given Lagrangian density \eqref{eqn:lagr}, the field equations for the baryons, leptons and the mesons can be derived, and given below,

\begin{eqnarray}
      (i\gamma^{\mu}\partial_{\mu}-m_{_B}^*-g_{\omega_ B} \gamma^{0}\omega_{0}-g_{\rho_B}\gamma^{0} I_{3_B}\rho_{03})\psi_{_B} &=&0 \\
      (i\gamma^{\mu}\partial_{\mu}-m_{\ell}){\psi}_{\ell} &=&0 
      \label{eqn:fieldeq}
\end{eqnarray}

\begin{eqnarray}
    m_{\sigma}^2\sigma &=& \sum_B g_{\sigma_B} n_B^s -\frac{\partial{U_{\sigma}}}{\partial \sigma} \\
    m_{\sigma^*}^2\sigma^* &=& \sum_B g_{\sigma^*_B} n_B^s \\
    m_{\phi}^2\phi_0 &=& \sum_B g_{\phi_B} n_B \\
    m_{\omega}^2\omega_0 &=& \sum_B g_{\omega_B} n_B-2\Lambda_{\omega} g_{\rho_N}^2 g_{\omega_N}^2 \rho_{03}^2 \omega_0 \\
       m_{\rho}^2\rho_{03} &=& \sum_B g_{\rho_B} I_{3_B}n_B-2\Lambda_{\omega} g_{\rho_N}^2 g_{\omega_N}^2 \omega_{0}^2 \rho_{03}
\end{eqnarray}
\label{eqn:mesonfieldeq}
where $n_B^s$ and $n_B$ are scalar and vector  baryon densities respectively.

\begin{eqnarray}
    n_B^s &=& \frac{g_{s_B}}{2\pi^2} \int_0^{k_{F_B}} \frac{m_B^* \  k^2}{\sqrt{k^2+{m_B^*}^2}}\  dk \nonumber \\
    n_B &=& \frac{g_{s_B} \  k_{F_B}^3}{6\pi^2} \nonumber
\end{eqnarray}
In the above expressions, `$g_{s_B}$' is spin degeneracy.  $k_{F_B}$ and `$I_{3_B}$'   are Fermi momentum and isospin projection of baryon `$B$' respectively. Expressions for field-dependent effective mass ($m^*_B$) and chemical potential ($\mu_B$) of baryon `$B$' are given below,
\begin{eqnarray}
    m^*_B &=& m_B-g_{\sigma_B}  \sigma - g_{\sigma^*_B}  \sigma ^* \nonumber\\
    \mu_B &=& \sqrt{k_{F_B}^ 2  + {m^*_B}^2 }  + g_{\omega_B} \omega_0 +g_{\phi_B} \phi_0 \nonumber \\
    &+& I_{3_B} g_{\rho_B} \rho_{03} 
    \label{eqn:effmass}
\end{eqnarray}

(NOTE: $g_{\sigma^*_B}$ = $g_{\phi_B}$ = 0 for nucleons.)
\\

In the NS core, baryons and leptons are in chemical equilibrium. Many works have described the procedure for solving such a composite system  (e.g., see  \cite{book1}). The system maintains chemical equilibrium via weak processes of the type  $B_1\rightarrow B_2 +\ell +\bar{\nu_{\ell}}$  and $B_2+\ell \rightarrow B_1 +{\nu_{\ell}}$ where $B$'s are baryons, $\ell$ represents  lepton and $\nu_{\ell}$ ($\bar{\nu_{\ell}}$) stands for  neutrino  (anti-neutrino). We focus on evolved neutron stars that are assumed to be transparent to neutrinos. Thus  neutrino chemical potential is set to `0', with this assumption all equilibrium conditions, may be summarized to a generic equation:

\begin{equation}
    \mu_i = b_i\mu_n-q_i\mu_e 
    \label{eqn:chemeq}
\end{equation}
Where $\mu_i$, $\mu_n$ and $\mu_e$ are chemical potentials of $i^{th}$ species, neutron and electron respectively, similarly $b_i$ and $q_i$ are baryonic  and electric charge of species `$i$' respectively.
\\

Charge neutrality of hadronic phase requires total charge should be `0', i.e :
\begin{equation}
    \sum_B q_{_B} n_{_B} -n_e-n_{\mu}=0 
    \label{eqn:chargeneut}
\end{equation}
where $q_{_B}$ is electric charge of baryon $B$ and $n_{_B}$, $n_e$, $n_{\mu}$ are number densities of baryon `$B$', electron and muon respectively.
\\

From the adopted Lagrangian one can construct the energy-momentum tensor, and obtain the energy density and pressure. The total energy density is given by :
\\
\begin{eqnarray}
    \epsilon &=&\frac{1}{2}m_{\sigma}^2\sigma^2+\frac{1}{2}m_{\sigma^*}^2{\sigma^*}^2+\frac{1}{2}m_{\omega}^2\omega_0^2+\frac{1}{2}m_{\rho}^2\rho_{03}^2\nonumber\\
    &+& \frac{1}{2}m_{\phi}^2\phi_0^2+U_{\sigma}+\sum_B \frac{g_{s_B}}{2\pi^2} \  
    \int_0^{k_{F_B}} {\sqrt{k^2+{m_B^*}^2}}\  dk \nonumber\\
    &+& 3 \Lambda_{\omega} (g_{\rho_N} g_{\omega_N} \rho_{03} \omega_0 )^2 \nonumber\\
    &+& \sum_{\ell} \frac{g_{s_{\ell}}}{2\pi^2} \ \int_0^{k_{F_{\ell}}} {\sqrt{k^2+{m_{\ell}}^2}}\  dk  \label{eqn:endens}
\end{eqnarray}
and pressure  ($p$) is given by the Gibbs-Duhem relation 
\\
 \begin{equation}
     p=\sum_{i=B,\ell} \mu_i n_i-\epsilon 
     \label{eqn:pres}
 \end{equation}
 
\subsection{Parameters of the model}
\label{subsec:para}

\subsubsection{Nucleon couplings}
To obtain the EoS, one must know the coupling constants, which are referred as the model parameters. As described in ~\Cref{sec:intro}, the nucleon isoscalar coupling constants  ($g_{\sigma_N},g_{\omega_N},b,c$) are obtained by fixing the nuclear saturation properties: nuclear saturation  density ($n_0$), binding energy per nucleon  ($E/A$), incompressibility  ($K$) and the effective nucleon mass  ($m^*$) at saturation. On the other hand, isovector coupling constants  ($g_{\rho_N},\Lambda_{\omega}$) are obtained by fixing the symmetry energy  ($J$) and the slope  ($L$) of symmetry energy at saturation  (see e.g., \cite{Hornick}). The saturation parameters considered in this work
have been summarized in \Cref{tab:rangepara}. \\

\begin{table}[h]
    \centering
\begin{tabular}
{|p{1.3 cm}|p{1cm}|p{1cm}|p{1cm}|p{1cm}|p{1cm}|p{1.5cm}|}
\hline
    Model & $n_0$ & $E_{sat}$& $K$& $J$ & $L$ & $m^*/m_N$  \\
      &($fm^{-3}$) & (MeV) & (MeV) & (MeV) & (MeV) & \\
\hline
    \hline
    
    RMF\cite{Hornick} &0.150&-16.0&240&32&60&[0.55-0.70]\\
    
    \hline
\end{tabular}
\caption{Range of saturation nuclear parameters used in this work.}   
\label{tab:rangepara}
\end{table}

Recently the effect of uncertainties in the saturation nuclear parameters on the $f$-mode frequencies was investigated in \cite{Sukrit}. It was shown that the effective nucleon mass $m^*$ has a significant effect while the other parameters have a negligible impact.
\\
\subsubsection{Hyperon couplings}

The vector-hyperon coupling constants  ($g_{\omega_Y},g_{\rho_Y},g_{\phi_Y}$) are fixed to their theoretical values from SU(6) quark model \cite{ChatterjeeNPA}.
\begin{eqnarray}
    g_{\omega_{\Lambda}}=g_{\omega_{\Sigma}}=2g_{\omega_{\Xi}}&=&\frac{2}{3}g_{\omega_N} \nonumber\\
g_{\rho_{N}} = g_{\rho_{\Xi}} &=& \frac{1}{2} g_{\rho_{\Sigma}} \> ; \> g_{\rho_{\Lambda}}=0 \nonumber\\
    2g_{\phi_{\Lambda}}= 2g_{\phi_{\Sigma}}=g_{\phi_{\Xi}}&=&\frac{-2\sqrt{2}}{3}g_{\omega_N}
\end{eqnarray}

The scalar-hyperon coupling constants are obtained by fixing the hyperon-nucleon  ($U_{Y}^{ (N)} (n_0)$) depth at normal nuclear matter \cite{ChatterjeeNPA}.
\begin{equation}
    U_{Y}^{N} (n_0)=-g_{\sigma_Y}\sigma+g_{\omega_Y}\omega_0
\end{equation}

Among the nucleon-hyperon potential depths, the best known is that of hyperon $\Lambda$ about $U_{\Lambda}^N$ = -30 MeV, and there are larger uncertainties for the $\Sigma$ and $\Xi$ potentials. State-of-the-art hypernuclear experiments indicate $\Sigma$ and $\Xi$ potential depths close to  +30 MeV and -18 MeV, respectively. 
\\

The hyperon-$\sigma^*$ coupling constants may be estimated by fixing the potential depth  ($U_{Y}^{Y^{\prime}}$), potential of a hyperon (Y) in a bath of the other hyperon  ($Y^{\prime}$) at normal nuclear matter density obtained from double hypernuclei data (\cite{Schaffner96},\cite{Schaffner93},\cite{Chatterjee06}). These are highly uncertain due to lack of sufficient experimental data \cite{Vidana}. Usually one adapts following values 
\begin{equation}
    U_{\Xi}^{\Xi}=2U_{\Lambda}^{\Lambda}=2U_{\Xi}^{\Lambda}=U_{\Lambda}^{\Xi}=-40\  MeV
\end{equation}

The effect of uncertainties in hyperon potential depths on maximum neutron star mass was studied systematically in \cite{ChatterjeeNPA}. It was also shown that the presence of $\sigma^*$ meson softens the hyperon EoS and renders it incompatible with current NS maximum masses. So in this work, we neglect the $\sigma^*$ meson and adapt the `$\sigma \omega \rho \phi$' model described in \cite{ChatterjeeNPA}. As for this work we do not consider attractive hyperon-hyperon interaction ($U_{Y}^{Y\prime}$) , from henceforth  $U_Y$  represents hyperon-nucleon ($U_Y^{(N)}$) potential depth.

 \subsection{EoS and Macroscopic description}
\label{subsec:eos}

 After calculating the coupling constants one can obtain the EoS (see e.g., \cite{book1} ). Given an EoS, 
 the macroscopic structure of neutron star can be obtained by solving the Tolman Oppenheimer Volkoff  (TOV) equations (e.g., see \cite{book1}) as given below;

\begin{eqnarray}
    \frac{dm}{dr}&=& 4\pi r^2 \epsilon (r) \label{eq:tov1} \\
    \frac{dp}{dr}&=&- \left[p (r)+\epsilon (r)\right]\frac{m (r)+4\pi r^3p (r)}{r (r-2m (r))} 
    \label{eq:tov2}
    \\
    p&=&p (\epsilon)  \label{eq:eos}
\end{eqnarray}
where Eqs.~(\ref{eq:tov1}, \ref{eq:tov2}) are equations of hydrostatic equilibrium and Eq. (\ref{eq:eos}) is the EoS.
\\

Integrating the TOV equations from the centre of the star to the surface, one can obtain global NS observables, such as mass ($M$), radius ($R$) and compactness ($C=M/R$). The tidal deformability ($\bar{\Lambda}$) can be obtained from the EoS by solving a set of differential equations coupled with the TOV equations \cite{YagiYunes2013}. For these calculations, we have included the crust EoS. However recent studies have confirmed that the crust may not significantly affect tidal polarizability \cite{PerotChamel2020,Tsang2019}. \\

\begin{figure}[htbp]
    \centering
    \includegraphics[width=\linewidth]{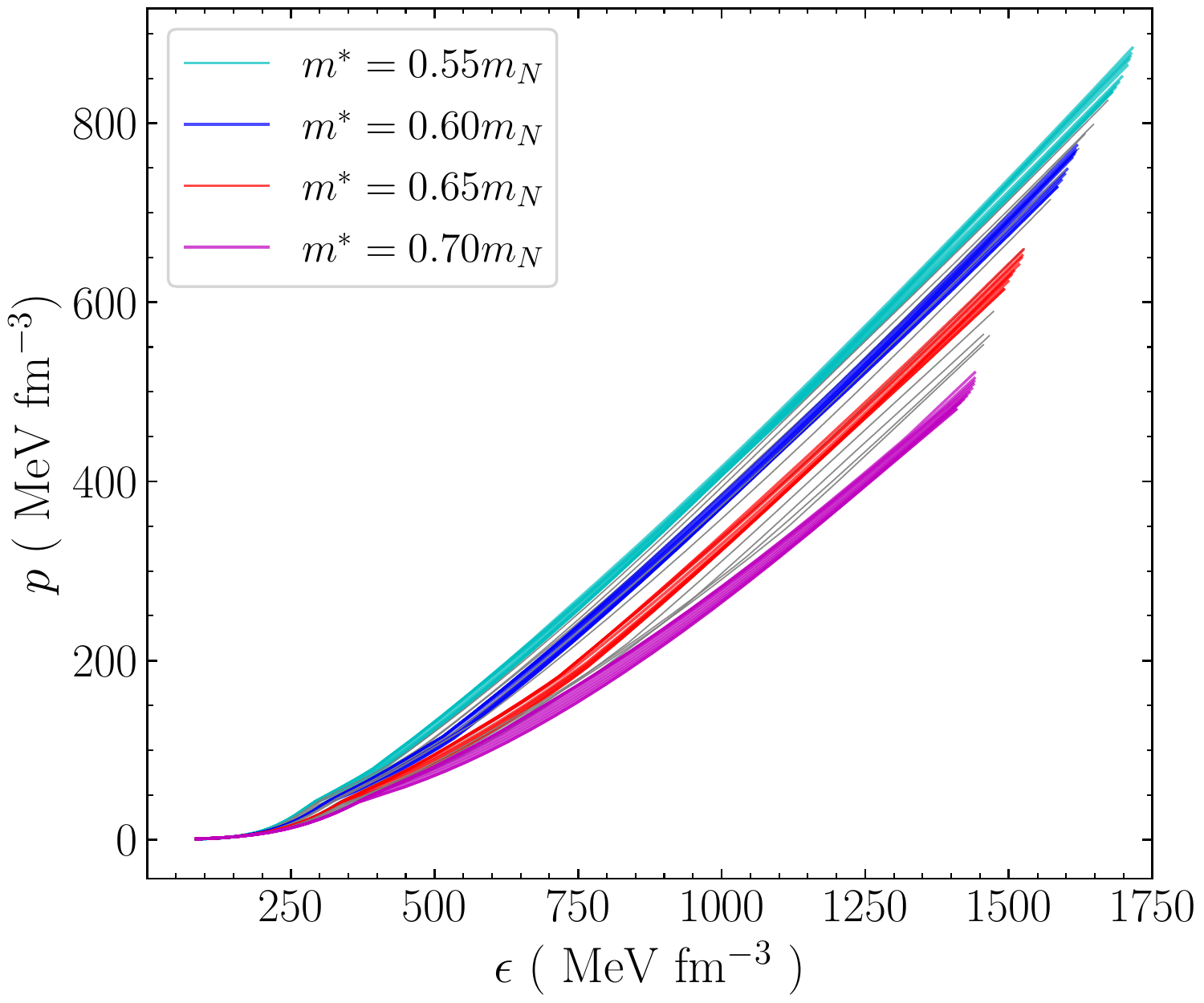}
    \caption{EoSs used in this work}
    \label{fig:alleos}
\end{figure}

We first tested our results using two commonly used parametrizations:  GM1 and GM3 \cite{GM1} for which the EoSs are well studied \cite{ChatterjeeNPA}. We considered hyperon potential depths $U_{\Lambda}=-30$ MeV, $U_{\Sigma}=+30$ MeV \cite{Millener88,Schaffner92,Mares95,SchaffnerGal00,FriedmanGal07} and $U_{\Xi}=-18$ MeV as suggested by the recent hypernuclear data \cite{Fukuda98,Khaustov00}. However, the potential depth of $\Xi$ hyperons is highly uncertain \cite{Vidana}. So in this work, we investigate the effect of uncertainty in `$U_{\Xi}$' along with that of the effective nucleon mass $m^*$ on $f$-mode oscillations. The EoSs considered are displayed in \Cref{fig:alleos}. 
\\

\begin{figure}[htbp]
    \centering
    \includegraphics[width=8.6cm]{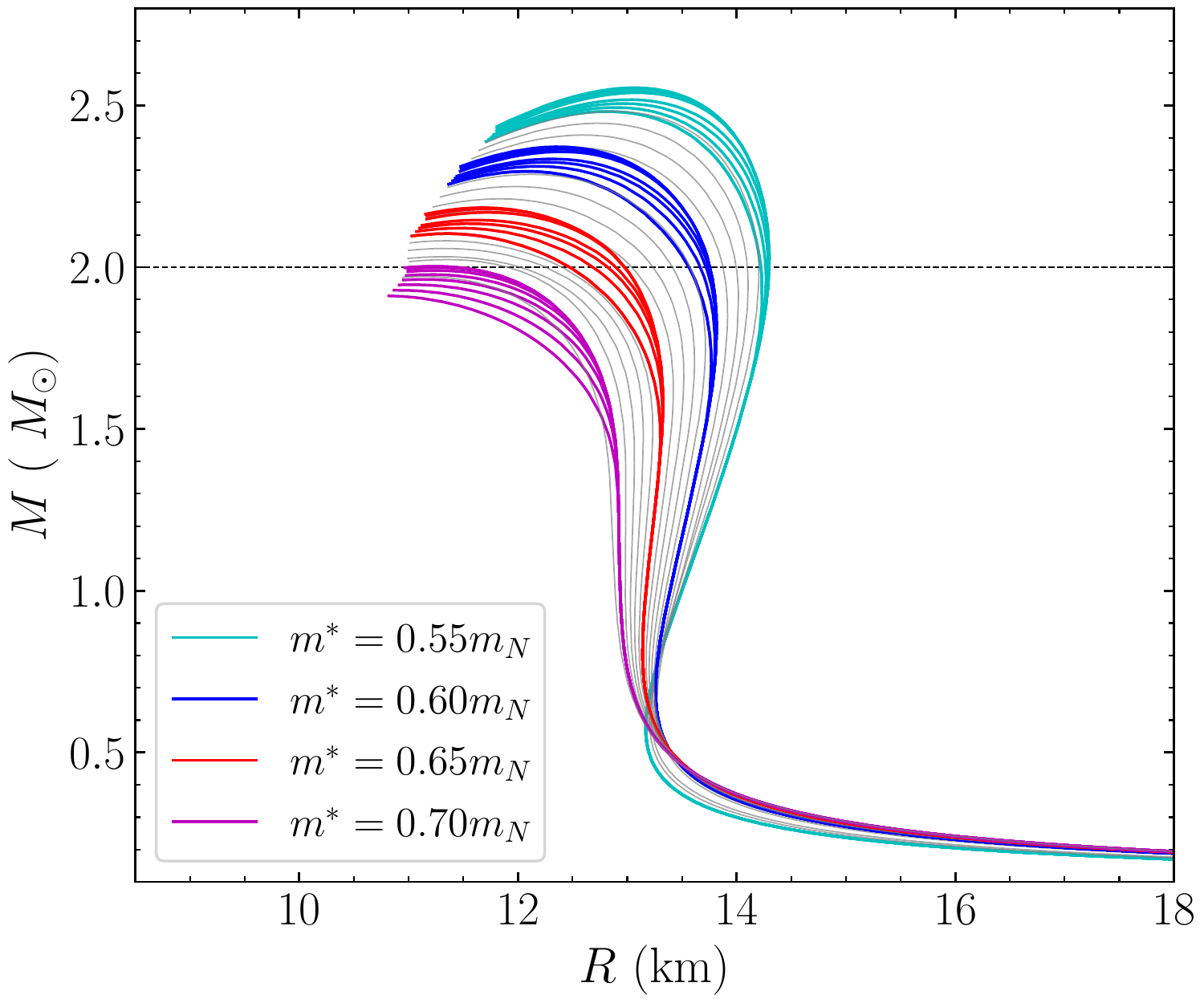}
    \caption{Mass-radius curves corresponding to EoSs in \Cref{fig:alleos}. The limit $2M_{\odot}$ is shown with a dashed line}
    \label{fig:allmr}
\end{figure}

\begin{figure}[htpb]
    \centering
    \includegraphics[width=\linewidth]{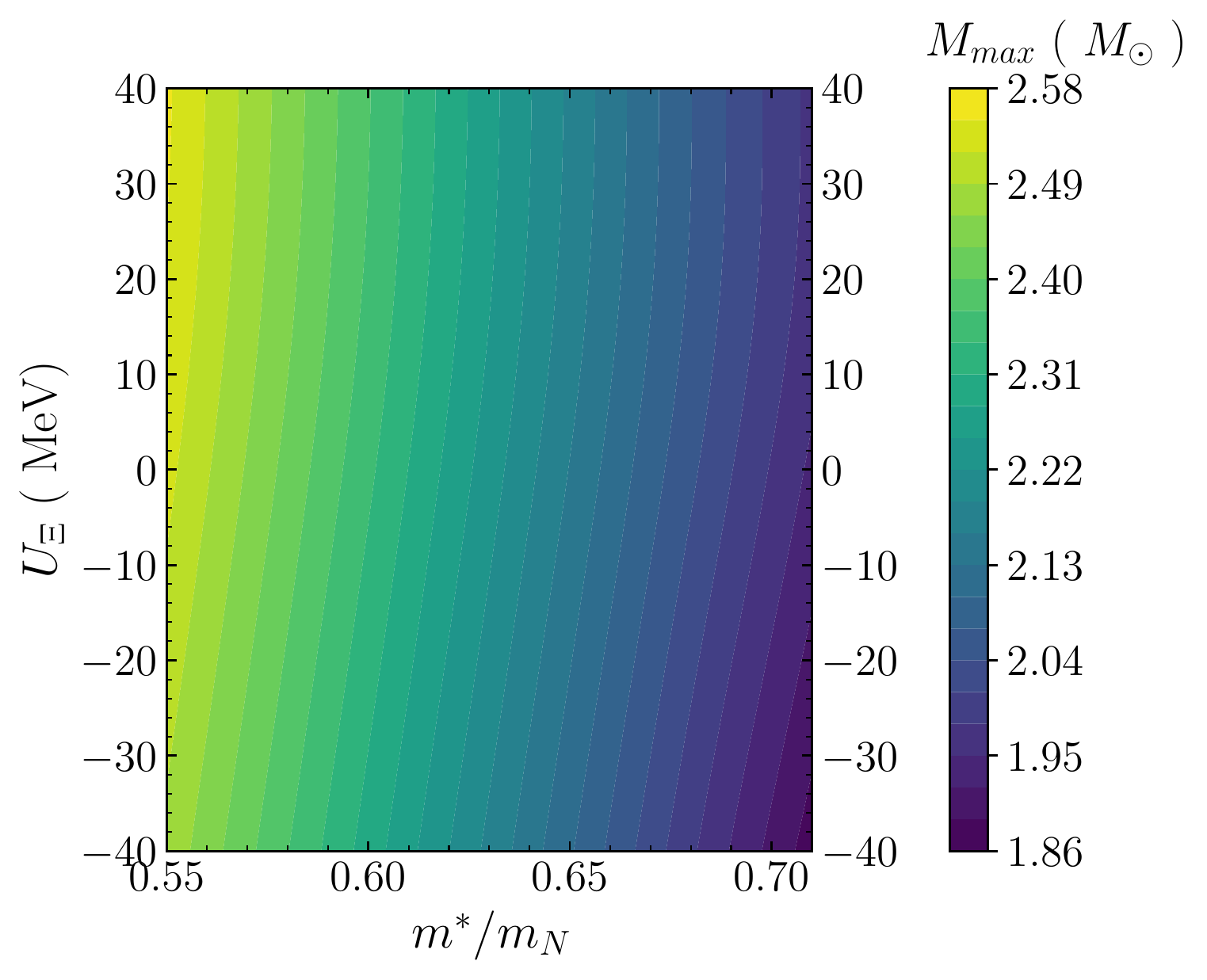}
    \caption{Variation of maximum mass in the $U_{\Xi}-m^*$ plane.}
    \label{fig:maxm_contour}
\end{figure}

In \Cref{fig:allmr} we display mass-radius (M-R) relations with parameters mentioned in \Cref{tab:rangepara}.
The effect of uncertainties in saturation parameters ( $m^*$ and $U_{\Xi}$) on maximum NS mass is displayed in \Cref{fig:maxm_contour}.
Imposing constraints from astrophysical observations would restrict the available parameter space: the 2$M_{\odot}$ constraint on the maximum mass \cite{Demorest2010, Antoniadis2013} is satisfied for nucleon effective mass $ \leq 0.7 \ m_N$ while the 2.1$M_{\odot}$ constraint \cite{Cromartie} is satisfied for $m^*/m_N \leq 0.67$. 
In this work, we will reckon with all those parameterizations which are compatible with the maximum 2$M_{\odot}$ constraint. 

\subsection{Calculation of oscillation modes}
\label{sec:modecalc}

NS oscillation modes have been studied for many decades. The procedure for non-radial oscillation modes in the non-relativistic framework was discussed by  Cowling \cite{Cowling} and in the general relativistic framework by Thorne and Campollataro \cite{Thorne}. In the vicinity of general relativity, one has to include the metric perturbation to solve the perturbed fluid equations. However, with the Cowling approximation in a weak gravitational field, one can neglect the metric perturbations. It was shown that the oscillation frequencies obtained for $f$-mode using Cowling approximation and including complete linearized equations of general relativity differ by less than 20\% \cite{Yoshida}.\\

 We want to investigate how the presence of hyperons affects the $f$-mode frequencies for non-rotating NSs. As explained above, we will work within Cowling approximation, so the spacetime metric for a spherically symmetric background is given by;
 \begin{equation}
     ds^2=-e^{2\Phi (r)}dt^2+e^{2\Lambda (r)}dr^2+r^2 d\theta^2+r^2\sin^2{\theta} d\phi^2 \label{eqn:metric}
 \end{equation}
 
In order to find mode frequencies one has to solve the following differential equations  (\cite{Sotani, Vasquez, Sandoval}):
\\
\begin{eqnarray}
    \frac{d W(r)}{dr}&=&\frac{d \epsilon}{dp}\left[\omega^2r^2e^{\Lambda (r)-2\phi (r)}V (r)+\frac{d \Phi(r)}{dr} W (r)\right] \nonumber \\
    &-& l(l+1)e^{\Lambda (r)}V (r) \nonumber \\
    \frac{d V(r)}{dr} &=& 2\frac{d\Phi (r)}{dr} V (r)-\frac{1}{r^2}e^{\Lambda (r)}W (r) 
    \label{eqn:perteq}
\end{eqnarray}
Where,
\begin{equation*}
    \frac{d \Phi(r)}{dr}=\frac{-1}{\epsilon(r)+p(r)}\frac{dp}{dr}
\end{equation*}
The functions $V (r)$ and $W (r)$ along with frequency $\omega$, characterize the Lagrange displacement vector  ($\zeta$) associate to perturbed fluid,
\begin{equation}
    \zeta^i=\left ( e^{-\Lambda (r)}W (r),-V (r)\partial_{\theta},-V (r) \sin^{-2}{\theta}\  \partial _{\phi}\right)r^{-2} Y_{lm} (\theta,\phi)
\end{equation}
where $Y_{lm} (\theta,\phi)$ is the  $lm$-spherical harmonic. Solution of Eq.~(\ref{eqn:perteq}) with the fixed background metric Eq.~(\ref{eqn:metric}) near origin will behave as follows:
\\
\begin{equation}
    W (r)=Ar^{l+1}, \> V (r)=-\frac{A}{l} r^l
\end{equation}
The vanishing  perturbed Lagrangian pressure at the surface will provide another constraint to be included while solving Eq.~\ref{eqn:perteq} which is given by,
\begin{equation} 
    \omega^2e^{\Lambda (R)-2\Phi (R)}V (R)+\frac{1}{R^2}\frac{d\Phi (r)}{dr}\Big|_{r=R}W (R)=0
\label{eqn:bc}
\end{equation}
Eqs.~(\ref{eqn:perteq}) are eigenvalue equations. Among the solutions, those that satisfy the boundary condition given by Eq.~(\ref{eqn:bc}) are the eigenfrequencies of the star. In this work, the solutions were obtained using the Ridder's method, which resulted in further improvement of our results as compared to the previous work in \cite{Sukrit}. In the mode calculations we included the effect of the NS crust and verified that inclusion of the crust has very little effect on the $f$-mode frequencies, compatible with the earlier investigations \cite{Flores2017}.
\\
\section{Results}
\label{sec:results}

\subsection{Uncertainty in effective mass}
\label{subsec:fmode_calculation}

\begin{figure}[htbp]
     \centering
     \includegraphics[width=\linewidth]{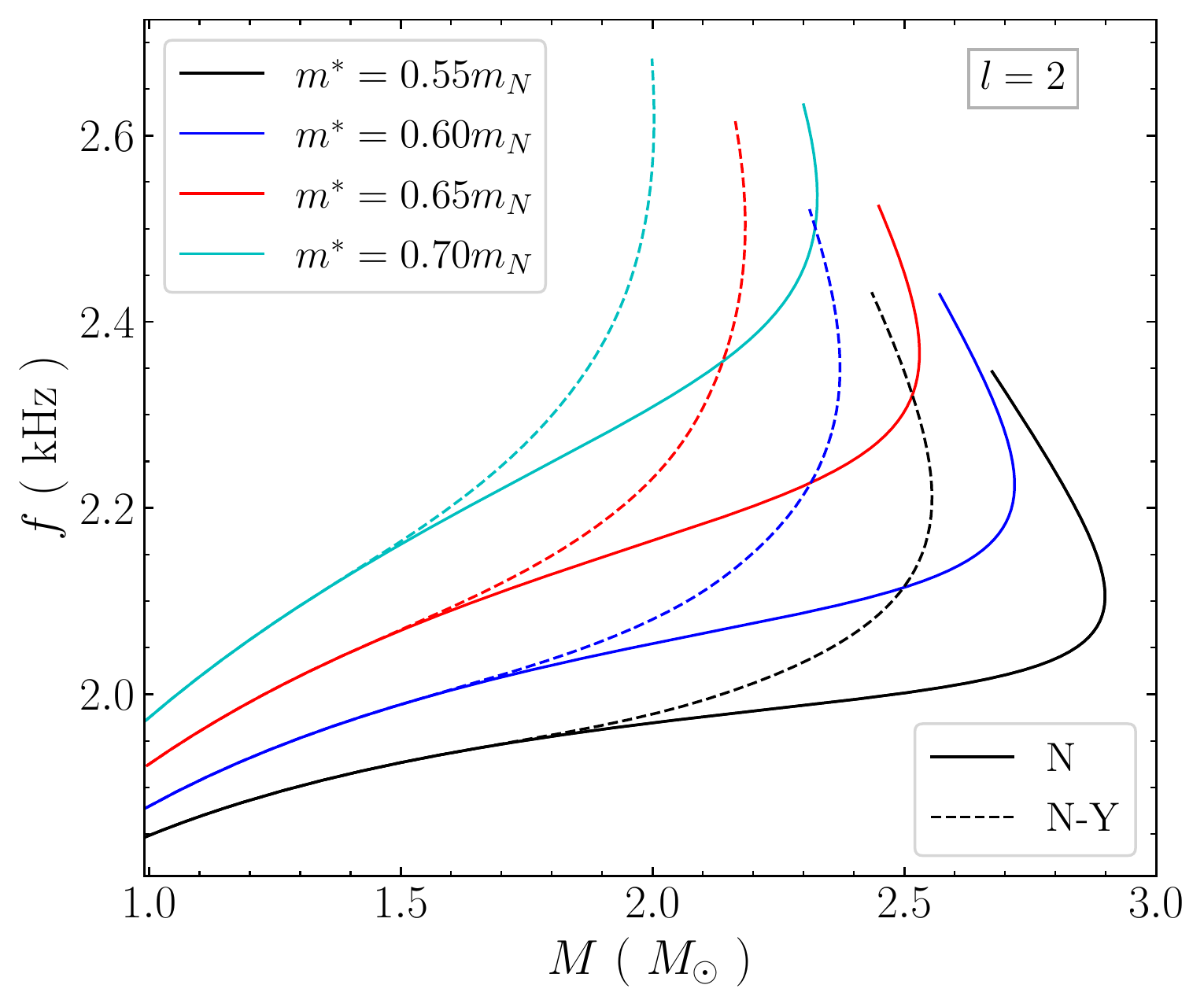}
     \caption{$f$-mode frequencies as a function of NS mass for different $m^*$ (i) for a nucleonic core (N)  (ii) for a hyperon-nucleon core (N-Y).}
     \label{fig:fmode_n_ny}
\end{figure}

In  \Cref{subsec:para},  we discussed the uncertainties associated with the nuclear saturation parameters and hyperon-potential depth, which in turn result in uncertainty in the  EoS. After testing our numerical scheme for $f$-mode frequencies, we reproduce the results from \cite{Sukrit} for a complete nucleonic core. As expected, we found that the effective nucleon mass has a significant effect, while the other saturation parameters have a negligible impact. We now extend our investigation, including hyperonic degrees of freedom. For clarity, only the variation of effective mass with and without hyperonic degrees of freedom is displayed in \Cref{fig:fmode_n_ny} for a fixed potential $U_{\Xi}$=-18 MeV. It is evident from the figure that inclusion of hyperons results in lower maximum masses and significantly higher $f$-mode frequencies. \\

\begin{figure}[htbp]
    \centering
    \includegraphics[width=\linewidth]{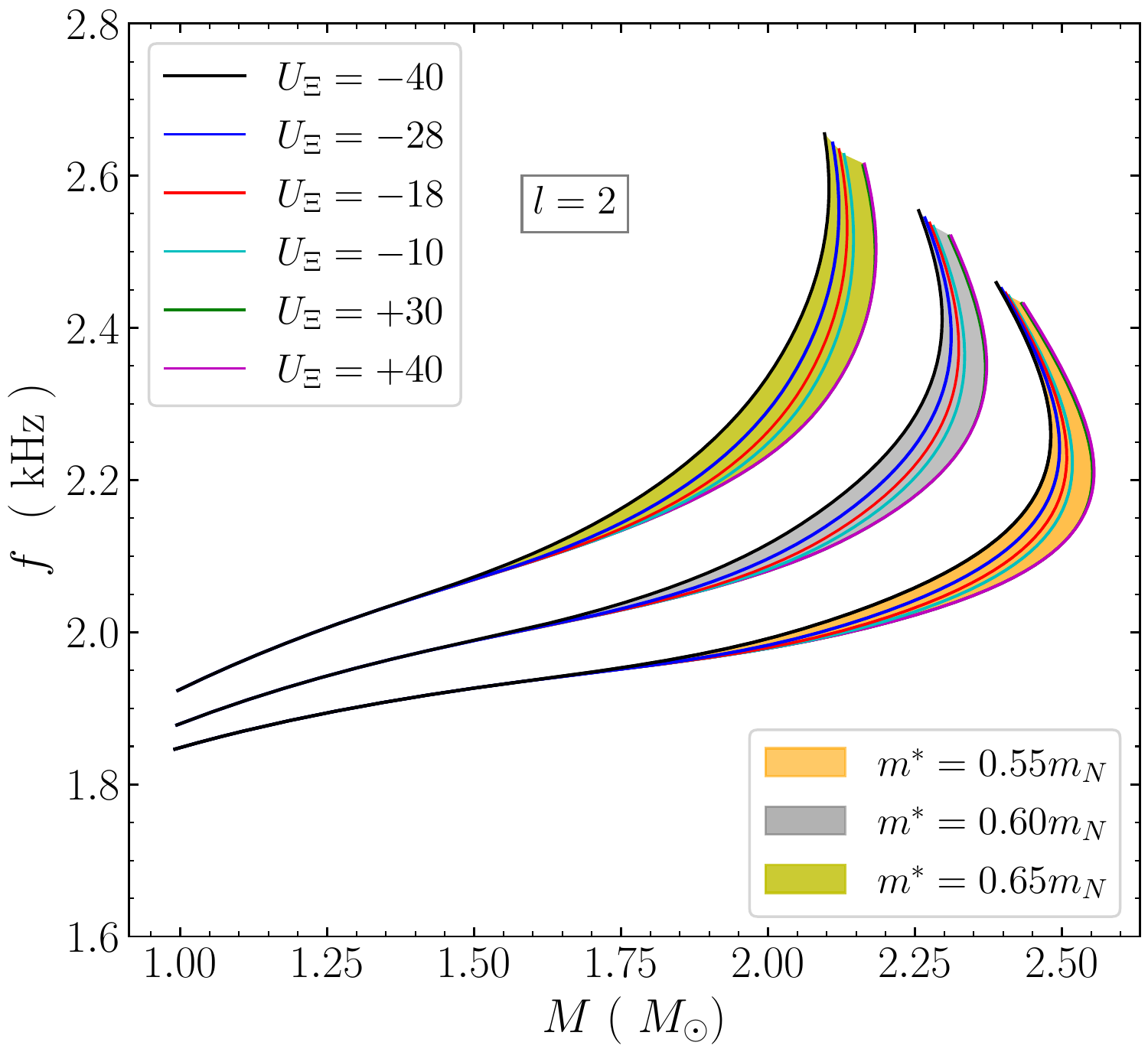}
    \caption{$f$-mode frequencies as a function of NS mass. Curves with same colour correspond to same $U_{\Xi}$ in MeV (e.g, all red curves correspond to $U_{\Xi}$=-18 MeV). Set of curves within a shaded area correspond to a fixed $m^*$ value (e.g, curves within yellow-shaded area are curves for different $U_{\Xi}$ but fixed $m^*=0.65\ m_N$) }
    \label{fig:l2fmode}
\end{figure} 

\begin{figure}[htbp]
    \centering
    \includegraphics[width=\linewidth]{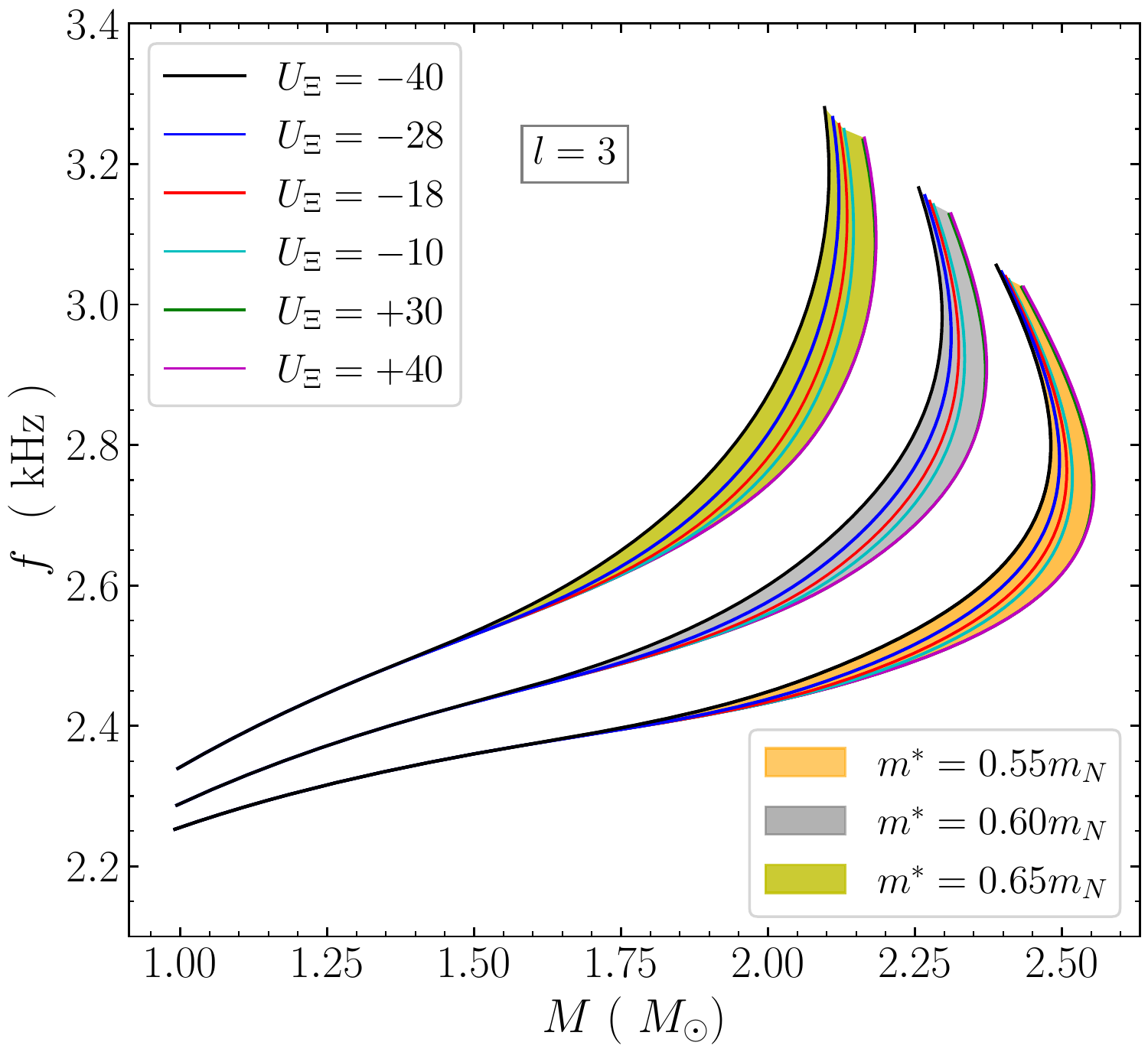}
    \caption{Same as \Cref{fig:l2fmode} but for higher order $l=3$ mode}
    \label{fig:l3fmode}
\end{figure}

\subsection{Uncertainty in hyperon potential}

 We further investigate the influence of uncertainties in both $m^*$ and $U_{\Xi}$ on the quadrupole ($l=2$) $f$-mode frequencies and present the result in  \Cref{fig:l2fmode}. From the figure, we conclude a non-negligible impact of variation in the $U_{\Xi}$ on $f$-mode frequencies for NSs with masses above $1.5 M_{\odot}$. The $f$-mode frequency changes over an interval of $\Delta_f\approx 0.09-0.15\ kHz$ for a range of $U_{\Xi}$ = -40 MeV to +40 MeV. \\

\begin{figure}[htbp]
    \centering
    \includegraphics[width=\linewidth]{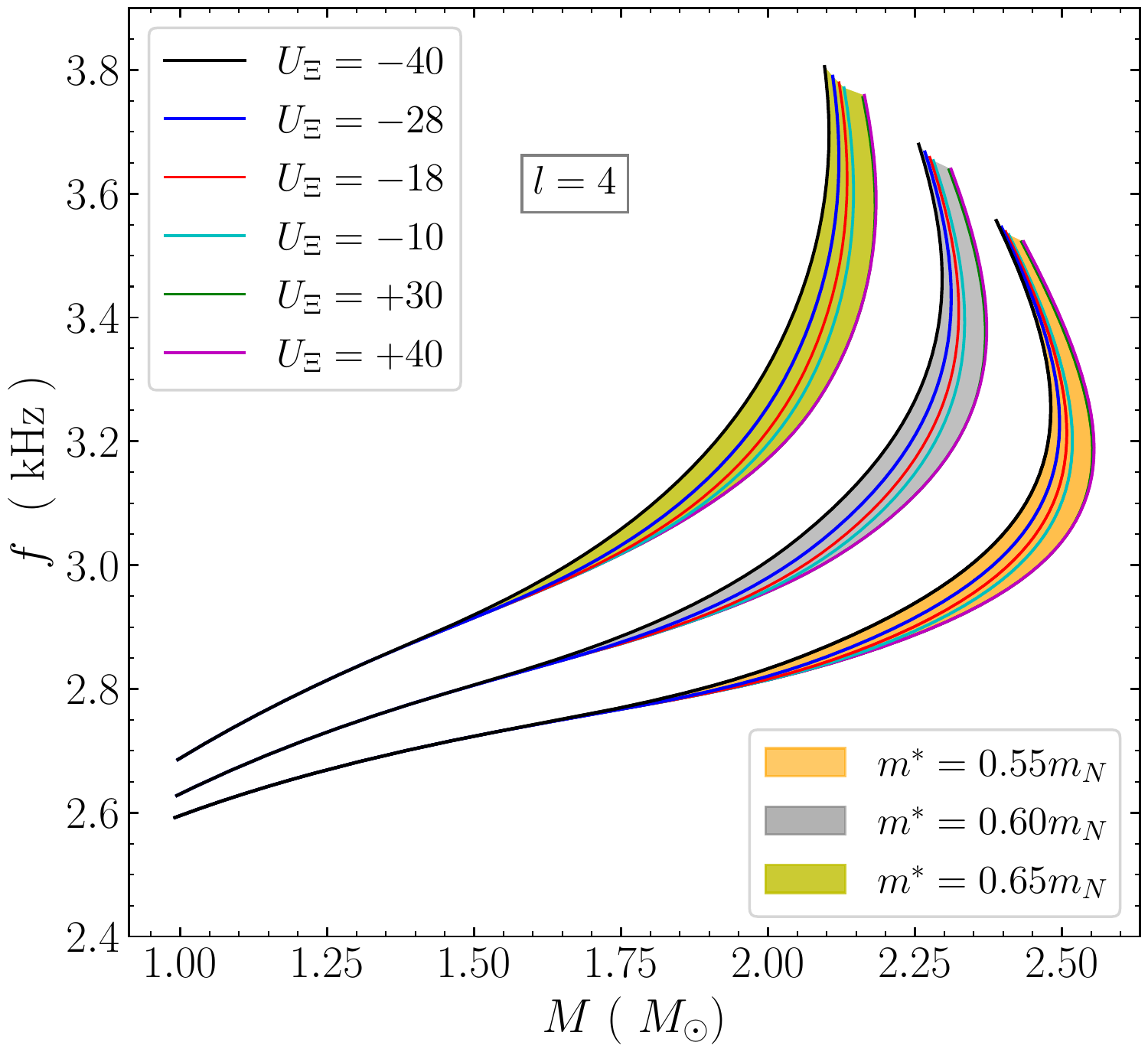}
    \caption{Same as \Cref{fig:l2fmode} but for higher order $l=4$ mode}
    \label{fig:l4fmode}
\end{figure}

However, the effective nucleon mass still has the most dominant influence on the $f$-mode frequencies. In \Cref{fig:l2fmode}, one can clearly differentiate between the curves with different effective mass in the $f$-modes as a function of the stellar mass. The frequency as a function of NS mass changes within an interval of 
1.8-2.6 kHz for the range of uncertainty in effective mass (0.55-0.65) compatible with 2$M_{\odot}$ maximum NS mass. The variation of $f$- mode frequencies is monotonically increasing with increasing $m^*$ and decreases with increasing $U_{\Xi}$ for a fixed $m^*$ (though smaller than the effect of $m^*$).
\\

A recent analysis of the instability window relevant to $f$-modes \cite{Kokkotas} in isolated NSs concludes that higher order $f$- modes ($l=3\ ,\ l=4\ etc.$) are more dominant than the quadrupole ($l=2$) $f$-modes. We therefore perform a similar investigation for higher order $f$-modes as done for quadrupole ($l=2$) $f$-modes. We present our results: (i) for $l=3$ in \Cref{fig:l3fmode}  and (iii) for $l=4$ in \Cref{fig:l4fmode}. The  higher order $f$-mode frequencies as a function of NS mass show qualitatively similar dependence on $m^*$ and $U_{\Xi}$ as $l=2$ $f$-mode frequencies. The frequency as a function of NS mass changes within an interval of 2.1-3.2 kHz and 2.5-3.8 kHz for $l=3$ and $l=4$ $f$-modes respectively. 

\subsection{$f$-modes and NS observables}
\label{subsec:fmodeandobservable}
\begin{figure}[htbp!]
     \centering
     \includegraphics[width=\linewidth]{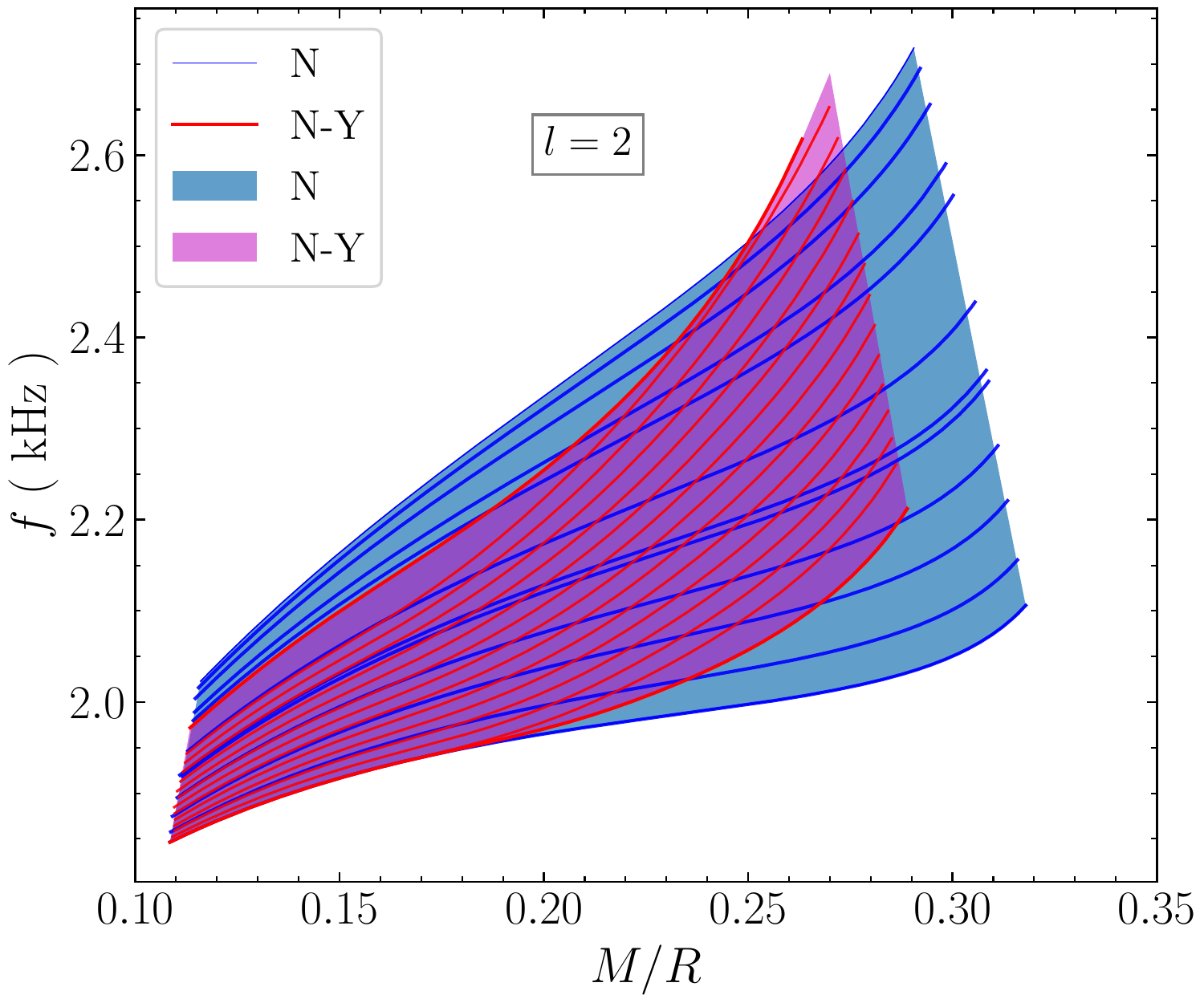}
     \caption{$f$-mode frequencies as a function of dimensionless stellar compactness ($M/R$). The range of frequencies spanned by nucleonic cores (shaded blue) and cores with nucleons and hyperons (shaded red) are presented. Different lines with same colour represent the same system ( nucleonic core (N) or nucleon-hyperon (N-Y)) with different parameterizations.}
     \label{fig:fmodecompactness}
 \end{figure}

In the previous section, we discussed the influence of the uncertainties in saturation parameters on $f$-mode frequencies as a function of stellar mass. To use GW frequencies for differentiating between different families of EoS, variations with the compactness of the star may be useful \cite{Andersson96,Andersson98, Benhar}. The compactness of a NS can be independently determined from the gravitational redshift derived from observations of spectral lines. In \Cref{fig:fmodecompactness}, we show the variation of $f$-mode frequencies as a function of stellar compactness.

\begin{figure}[htbp!]
    \centering
    \includegraphics[width=\linewidth]{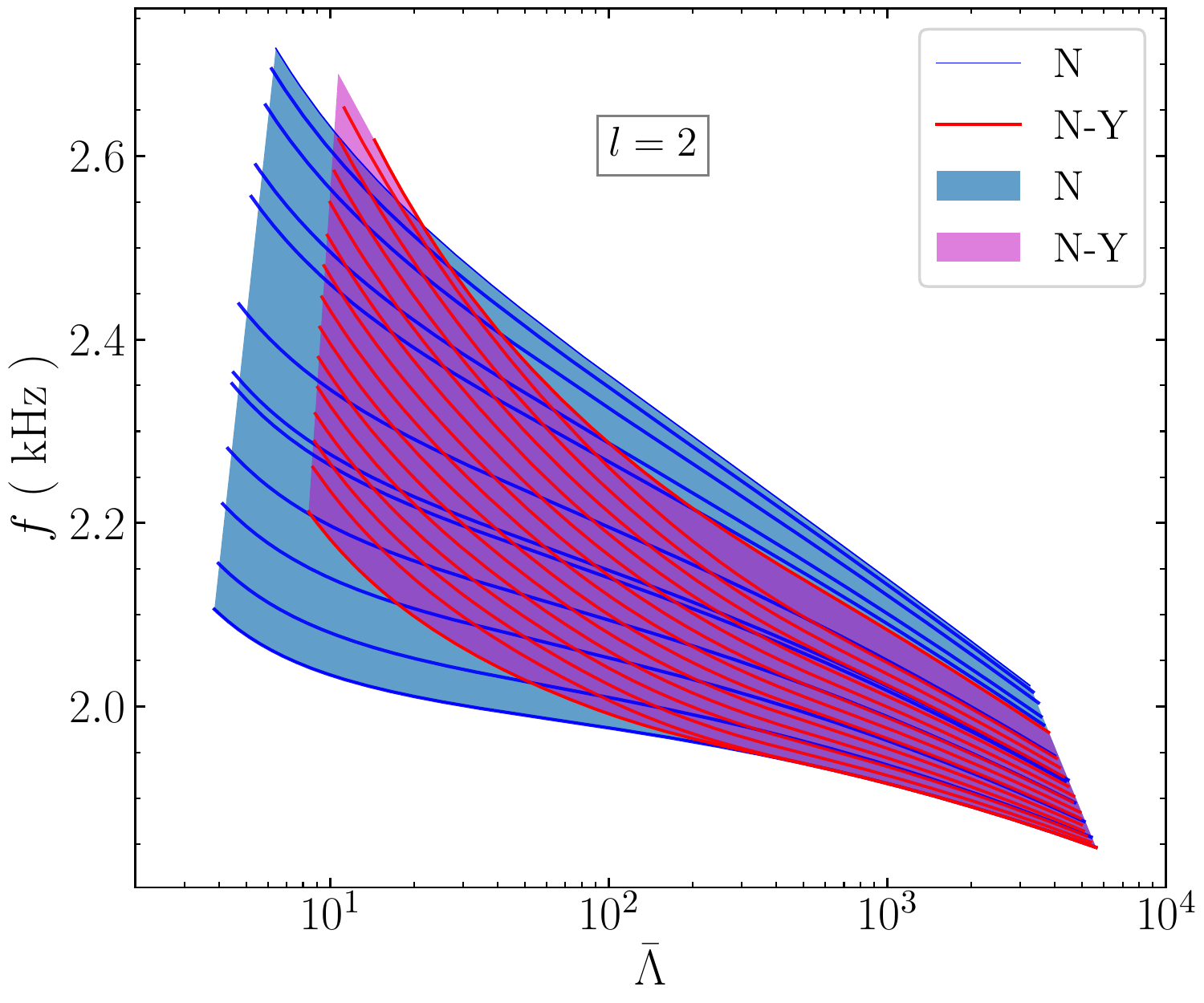}
    \caption{$f$-mode frequencies as a function of dimensionless tidal deformability ($\bar{\Lambda}$). The range of frequencies spanned by nucleonic cores (shaded blue) and cores with nucleons and hyperons (shaded red) are presented. Different lines with same color represents to same system ( nucleonic core (N) or nucleon-hyperon (N-Y)) with different parameterizations.}
    \label{fig:fmodetidal}
\end{figure}

Along with global observables like mass, radius, and compactness, another important observable is tidal deformability, which plays an important role in constraining the underlying EoS. In the inspiral phase, the NSs exert strong gravitational forces on each other, and the deformation produced depends on their EoSs (\cite{Hinderer, Prakash}). Measuring both $f$-mode frequency and tidal deformability independently would then allow us to constrain the NS composition. We present the variation of $f$-mode frequencies as a function of dimensionless tidal deformability ($\bar{\Lambda}$) in \Cref{fig:fmodetidal}. We found our lower limit of $f$-mode frequency is in good agreement with the limit obtained from observations of GW170817 \cite{Pratten}. 

\section{GW Asteroseismology}
\label{sec:asteroseismology}

The motivation behind the neutron star GW `asteroseismology' is similar to traditional helioseismology: to use GW observations of pulsating neutron stars as probes of their interiors. In NS asteroseismology the key idea is to parametrize an oscillation mode's angular frequency and GW damping timescale in terms of the NS's bulk parameters like mass, radius, and the angular frequency. Inverse asteroseismology allows us to construct EoS-insensitive relations by making a combination of GW observational and global parameters of NS like mass, radius, and rotational frequency (for rotating stars).
The idea of GW asteroseismology was first put forward by \cite{Andersson96} for polytropic EoSs and then investigated in ``AK" \cite{Andersson98} with selective realistic EoSs. Many of the selective EoSs considered in this work are now rendered incompatible with the maximum mass $2M_{\odot}$ observational constraint, and are ruled out. AK derived an empirical asteroseismology relation for $f$-mode frequencies as a function of average density,
\begin{equation}
 f(\text{kHz}) = a + b \sqrt{ \frac{\bar{M}}{{\bar{R}^3}} } 
 \label{eq:fit}
\end{equation}
in terms of dimensionless parameters $\bar{M}=\frac{M}{1.4  M_{\odot}}$ and $\bar{R}=\frac{R}{10 \text{km}}$ (Eq.~5 of their paper~\cite{Andersson98}).\\

This was further probed by \cite{Benhar} using chosen EoSs including exotic matter (hyperons and quarks). It was shown (in Figure 2 of their paper~\cite{Benhar}) that the previous ``AK fit" was now modified by the inclusion of a wider range of EoSs. Later work by \cite{Doneva} derived asteroseismology relations for rotating neutron stars with 5 chosen realistic EoSs, and provided new fits derived from their non-rotating limit (Eq.~25-27 of~\cite{Doneva}). A few chosen EoSs including exotic forms of matter (hyperons and quarks) have also been considered in some recent works \cite{Vasquez, Salcedo}. The different values $a$ and $b$ for the fit relation (Eq.~\eqref{eq:fit}) for these works are summarized in  \Cref{tab:fitrelations}. A comparison of the asteroseismology relations from these studies shows that the results are sufficiently different due to the difference in the choice of EoSs. \\

 The knowledge of mode frequencies and the NS mass (which is among the most precisely determined global variables) can help to discriminate among the different EoSs. The empirical relations could then be used to determine the mass and radius of the NS therefore its EoS: its stiffness, the presence of hyperons/quarks \cite{Benhar}. \\

\subsection{Fit relations}
\label{subsec:fitrelations}
In this work, instead of taking a few selected EoSs, we aim to investigate the effect of full range of uncertainties in nuclear and hypernuclear saturation parameters on the EoS and asteroseismology relations, subject to constraints from astrophysical observations. Solving the pulsation equations within a full linearized GR framework results in complex solutions, of which the real part can be interpreted as frequency and the imaginary part as damping time. However, within this work, we used the Cowling approximation for obtaining the mode frequencies, which results in only real solutions (frequency only).\\

\begin{figure}[htbp]
    \centering
    \includegraphics[width=\linewidth]{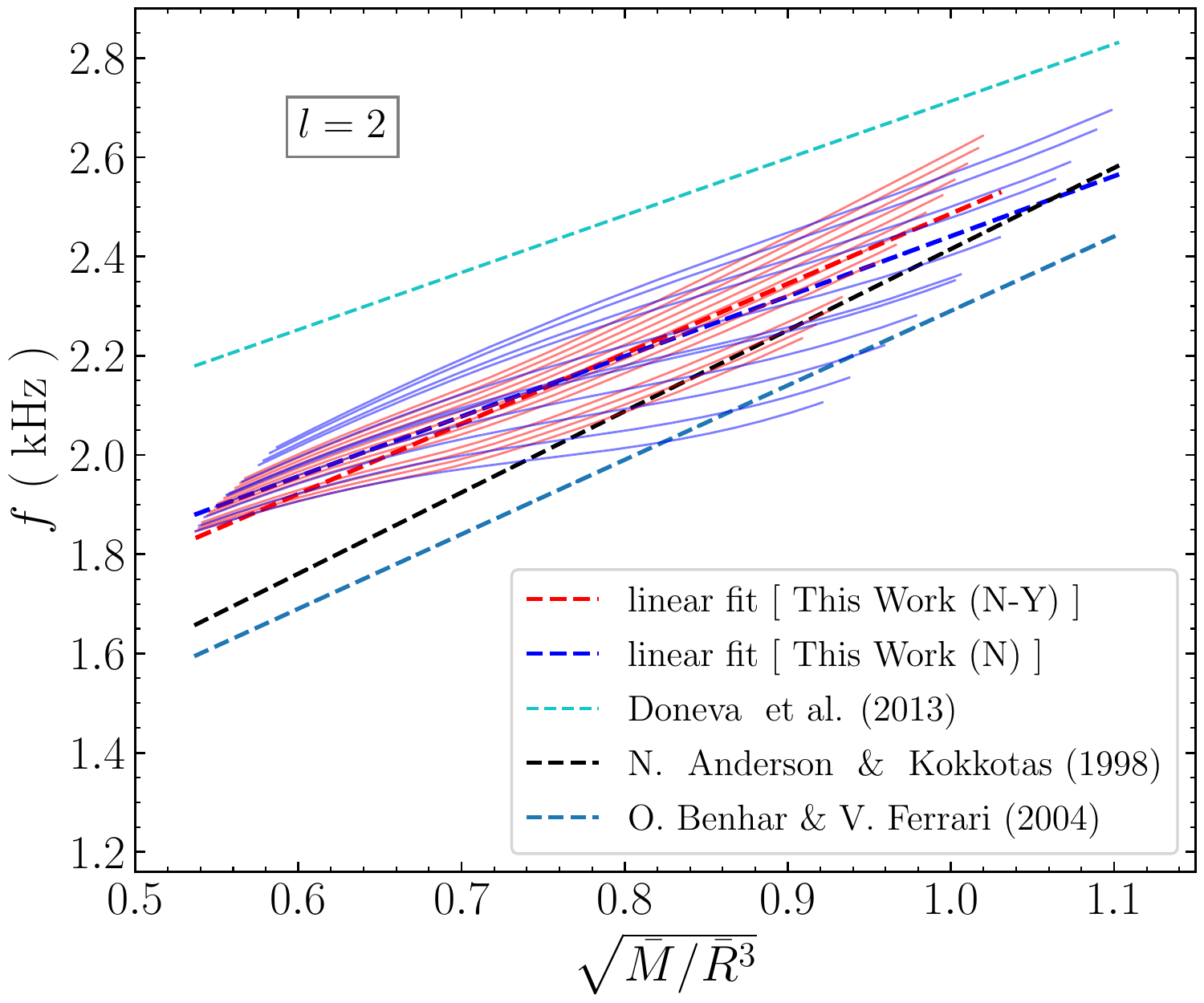}
    \caption{Mode frequencies for $l=2$ mode as a function of the average density of the star, for variation in both $m^*$ and $U_{\Xi}$. Red solid lines  correspond to stars with nucleon- hyperon core (N-Y) and blue solid lines to stars with only nucleonic (N) core. Linear fit relations from this work and other works (\Cref{tab:fitrelations}) are shown with dashed lines.}
    \label{fig:l2linearfit}
\end{figure}

We derive the empirical asteroseismology relation for $f$-mode frequencies as a function of average density (Eq.~\ref{eq:fit}). In terms of dimensionless parameters $\bar{M}=\frac{M}{1.4  M_{\odot}}$ and $\bar{R}=\frac{R}{10\ km}$, the fit relation for $l=2$ mode is given in Eq.~\eqref{eq:l2fitrelation}. We display the $f$-mode frequencies as a function of $\sqrt{\bar{M}/ \bar{R}^3}$ along with  asteroseismology relations from previous works for $l=2$ mode in \Cref{fig:l2linearfit} and compare the obtained fit relation with previous works in \Cref{tab:fitrelations}.  The fit relation in \eqref{eq:l2fitrelation} is obtained by varying both parameters (effective nucleon mass and hyperon potentials). From Eq.~\eqref{eq:l2fitrelation}, one can quickly see that for a canonical NS of ${1.4  M_{\odot}}$ and 10 km radius, the $f$-mode frequency is 2.49 kHz.

\begin{table}[h]
   \centering
\begin{tabular}{|p{5.3cm}|p{1.1cm}|p{1.1cm}|}
\hline
    Works   & $a$ (kHz) & $b$ (kHz)  \\
\hline
\small{N.Andersson and Kokkotas(1998) \cite{Andersson98}} & 0.78 & 1.635\\
\hline
 D.Doneva et al. (2013) \cite{Doneva} &1.562& 1.151\\
 \hline
   O.  Benhar and V.  Ferrari (2004) \cite{Benhar} & 0.79 & 1.500\\
  \hline
  This work (N-Y) \eqref{eq:l2fitrelation} & 1.075 & 1.412\\
  \hline
\end{tabular}
\caption{Asteroseismology relations from different works. Where `$a$' and `$b$' are related to quadrupole ($l=2$)  f-mode frequency by: $f \ (\text{kHz})=a+b\sqrt{\frac{\bar{M}}{\vbox{\kern.3ex\hbox{$\scriptscriptstyle\bar R^3$}}}}$}
\label{tab:fitrelations}
\end{table}

\begin{equation}\label{eq:l2fitrelation}
    f(\text{kHz})=1.075+1.412\sqrt{\frac{\bar{M}}{\bar{R}^3}}
\end{equation}


\subsubsection{Higher order f-modes}
\label{subsec:fithigherordermodes}

\begin{figure}[htpb]
    \centering
     \includegraphics[width=\linewidth]{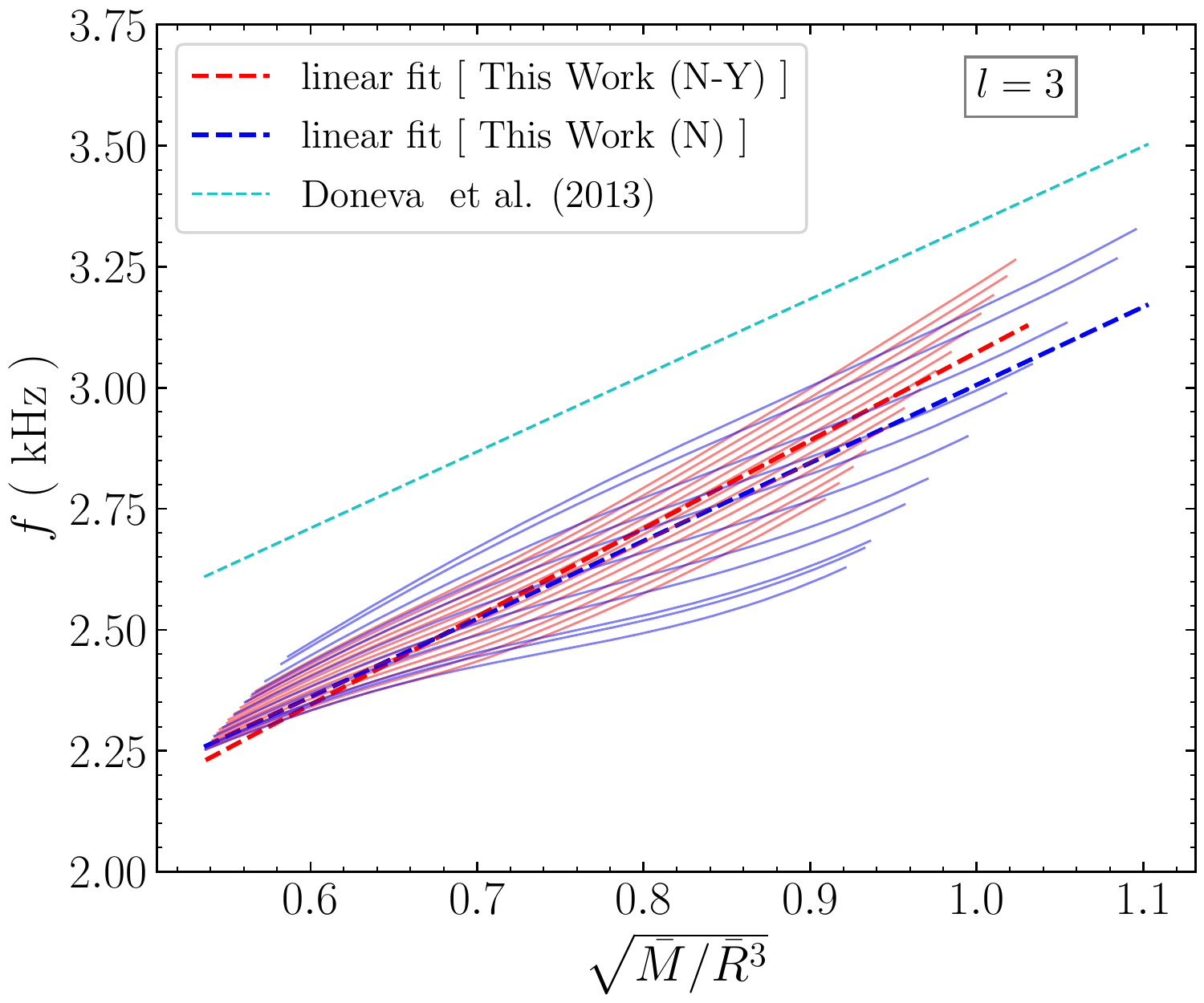}
    \caption{Same as \Cref{fig:l2linearfit} but for $l=3$ $f$-mode frequencies.}
    \label{fig:l3linearfit}
\end{figure}

\begin{figure}[htpb]
    \centering
     \includegraphics[width=\linewidth]{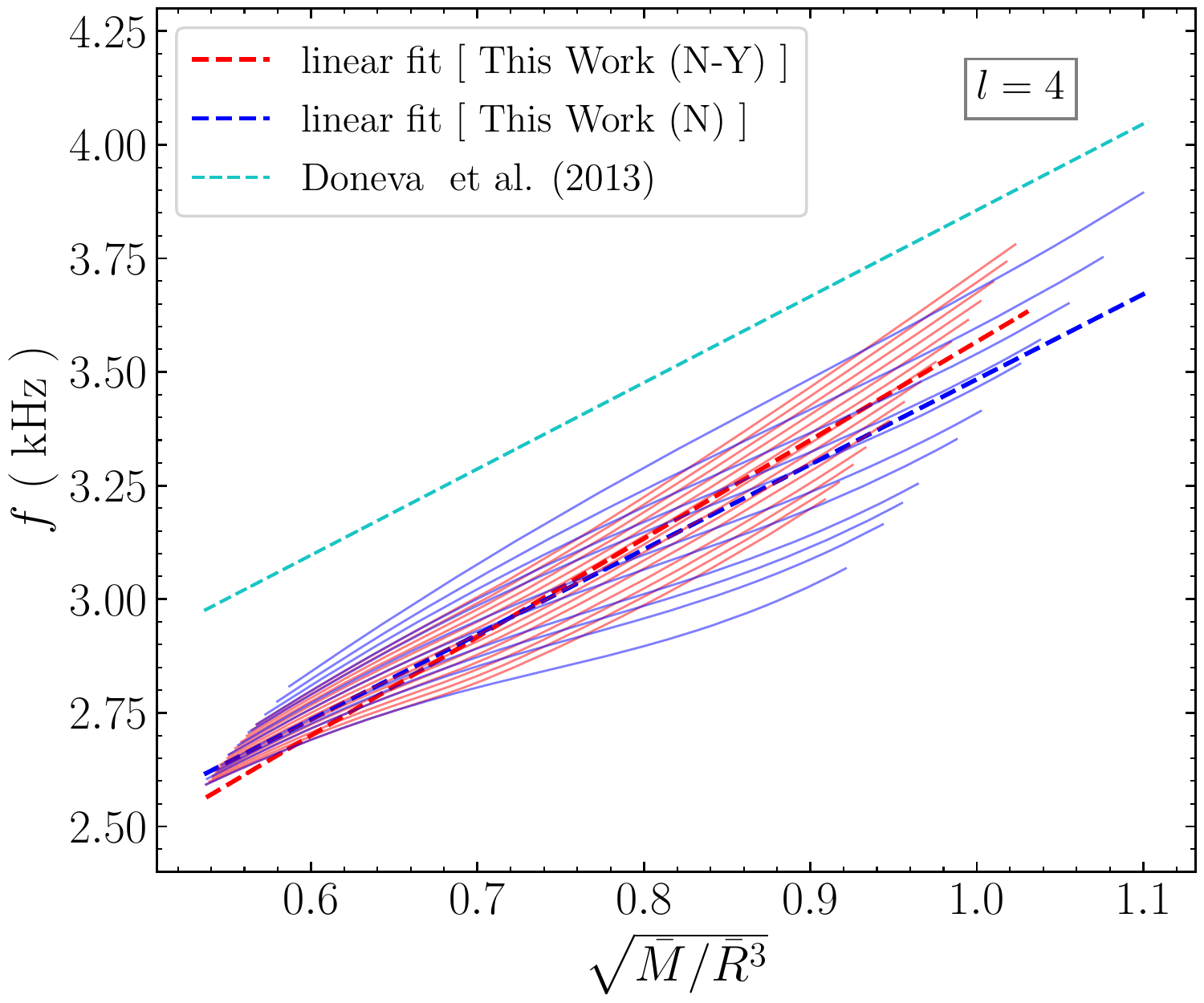}
    \caption{Same as \Cref{fig:l2linearfit} but for $l=4$ $f$-mode frequencies.}
    \label{fig:l4linearfit}
\end{figure}

Recent investigations \cite{Kokkotas} conclude higher order $f$-modes in isolated NSs could be more dominant over quadrupole $f$-mode. The asteroseismology relations for higher order $f$-modes was carried out by \cite{Doneva} with few selective EOSs. They also conclude that observation of two $f$-modes can be used to determine mass, radius to a good accuracy. So, we extend our investigation for next higher order $f$-modes ($l=3\ \text{and}\ l=4 $). We present our results for $l=3$ and $l=4$ $f$-mode frequencies as a function of $\sqrt{\bar{M}/ \bar{R}^3}$ along with relations from previous work \cite{Doneva} in \Cref{fig:l3linearfit} and \Cref{fig:l4linearfit} respectively. The asteroseismology relations in terms of dimensionless parameters ($\bar{M},\ \bar{R}$) for $l=3$ and $l=4$ are given in Eq.~\eqref{eq:l3fitrelation} and Eq.~\eqref{eq:l4fitrelation} respectively.

\begin{equation}\label{eq:l3fitrelation}
    f(\text{kHz})=1.254+1.812\sqrt{\frac{\bar{M}}{\bar{R}^3}}
\end{equation}
\begin{equation}\label{eq:l4fitrelation}
    f(\text{kHz})=1.401+2.167\sqrt{\frac{\bar{M}}{\bar{R}^3}}
\end{equation}

We find that unlike previous existing investigations of $f$-modes with hyperonic EoSs, the difference between nucleonic and hyperonic fit relations is quite small when the entire space of uncertainties is taken into account. Our fit relations vary widely from those who have considered selected EoSs in their study.

\subsection{Universal Relations}
\label{subsec:universalrelations}
\begin{figure}[htbp]
    \centering
    \includegraphics[width=\linewidth]{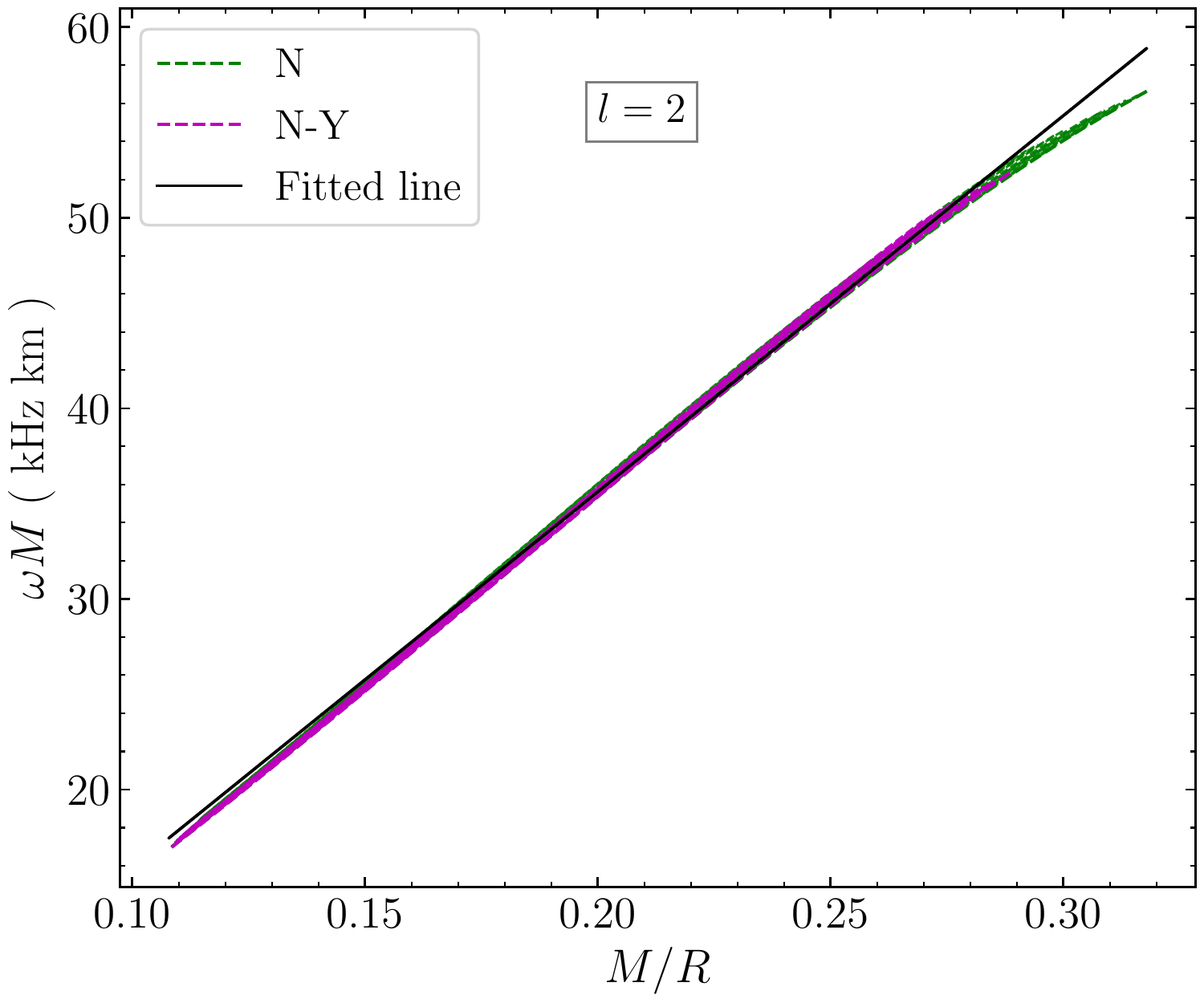}
    \caption{Angular frequencies ($\omega=2\pi f$) scaled by mass ($\omega M$) as a function of stellar compactness.}
    \label{fig:omega_m}
\end{figure}

\begin{figure}[htbp]
    \centering
    \includegraphics[width=\linewidth]{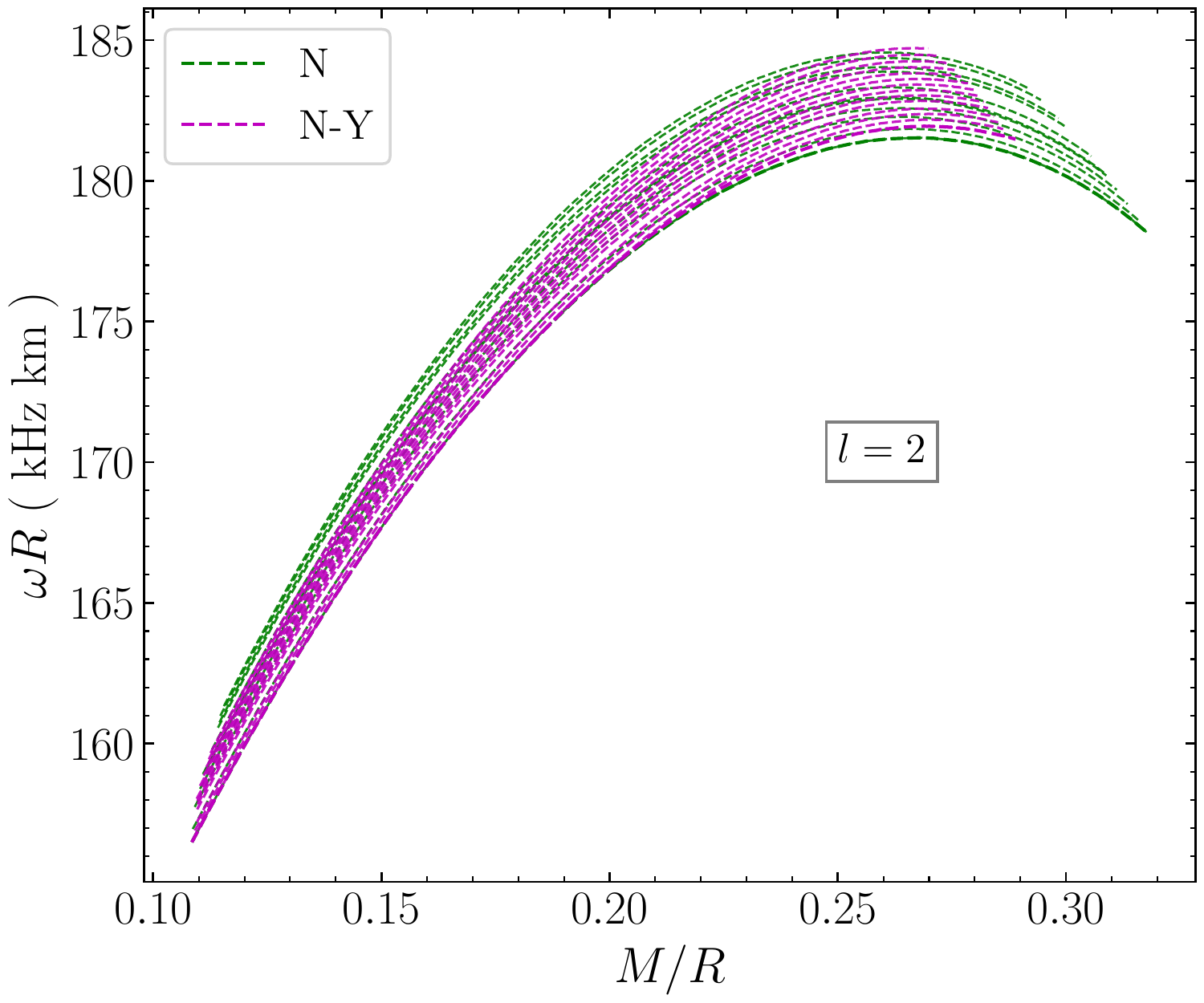}
    \caption{Angular frequencies ($\omega=2\pi f$) scaled by radius ($\omega R$) as a function of stellar compactness.}
    \label{fig:omega_r}
\end{figure}

Correlations between different mode frequencies scaled appropriately by NS's mass or radius with stellar compactness was first  suggested by \cite{Andersson96}. These phenomenological relations are quite independent of underlying matter composition of the NS. It was also suggested that, detection of  mode frequencies along with these hypotheses could help us to determine the NS's global properties. These hypotheses were studied for $g$ modes in \cite{Sotani}, for $p$, and $w$ modes in \cite{Salcedo} and for $f$-modes in \cite{Wen}. We test these hypotheses for the EoSs considered in this work and present the variation of angular frequency  scaled by mass (radius) as a function of stellar compactness in \Cref{fig:omega_m} (\Cref{fig:omega_r}). The scaled frequency with mass ( $\omega M$) behaves with compactness  universally, while  ($\omega R$) deviates slightly. 
We find that the $f$-mode frequency can be well expressed as a function of dimensionless stellar compactness $M/R$ using the fit relation \eqref{eq:universal}.
\begin{equation}\label{eq:universal}
\omega M \ (\text{kHz km})= 197.295 \left(\frac{M}{R}\right)-3.836.
\end{equation}


 \section{Conclusions}
 \label{sec:conclusions}
 
Previous works on $f$-modes suggested the inversion method in asteroseismology, by proposing fit relations between $f$-mode frequencies and global observables such as mass, density or compactness. For such works, selected EoSs available at that time were considered, many of which have now been rendered incompatible with large NS mass observations. Hence the fit relations based on these works are not applicable anymore, as the fit formulas strongly depend on the chosen EoSs. There are very few investigations in the literature extending the $f$-mode calculations to the hyperonic sector \cite{Benhar, Salcedo}. In \cite{Salcedo}, the hyperon EoSs used (WCS1 and WCS2) were two specific parametrizations chosen out of the full parameter study performed by one of our authors (D.C.) \cite{ChatterjeePRC}. Not imposing such universal relations opens up the possibility of understanding the composition and phase transitions in dense matter.
\\

In this work, for the first time we studied the influence of the uncertainties in the underlying nuclear and hypernuclear physics on $f$-mode frequencies. Within the RMF model framework, we performed a systematic investigation of the impact of the uncertainties in the dominant parameters, the nucleon effective mass and hyperon potential depth, consistent with recent nuclear experimental data, on the $f$-mode frequencies. The EoSs considered in this work are compatible with $2M_{\odot}$ maximum mass limit. The presence of hyperonic degrees of freedom affects the $f$-mode frequencies for NSs with mass above 1.5 $M_{\odot}$, and the $f$-mode frequencies increase with respect to the frequencies of a pure nucleonic core.\\

We conclude that nucleon effective mass has a significant effect on mode frequencies, whereas potential depth has a non-negligible impact on mode frequencies. 
For the complete range of uncertainty associated with saturation parameters and imposing maximum  $2M_{\odot}$ constraint $f$-mode frequencies vary between 1.8-2.6kHz, 2.2-3.3 kHz and 2.5-3.8 kHz for $l=2$, $l=3$ and $l=4$ respectively. 
For the change of $\Xi-N$ potential depth within  a range -40 to +40, $f$-mode frequencies change within a range of 0.09-0.15 kHz. 
\\

An important motivation for this work is to try and distinguish between the EoS with and without hyperons using observations of $f$-mode frequencies. In previous works \cite{Benhar}, it was concluded that the transition to hyperon matter produces significant softening of the EoS leading to lower maximum masses, and the corresponding $f$-mode frequencies would be considerably higher than the nucleonic case. One may then distinguish the internal composition using frequency vs mass curves. In \Cref{fig:fmode_n_ny}, we obtained higher frequencies and lower masses in presence of hyperons as expected. 
\\

\begin{figure}[htbp]
    \centering
    \includegraphics[width=\linewidth]{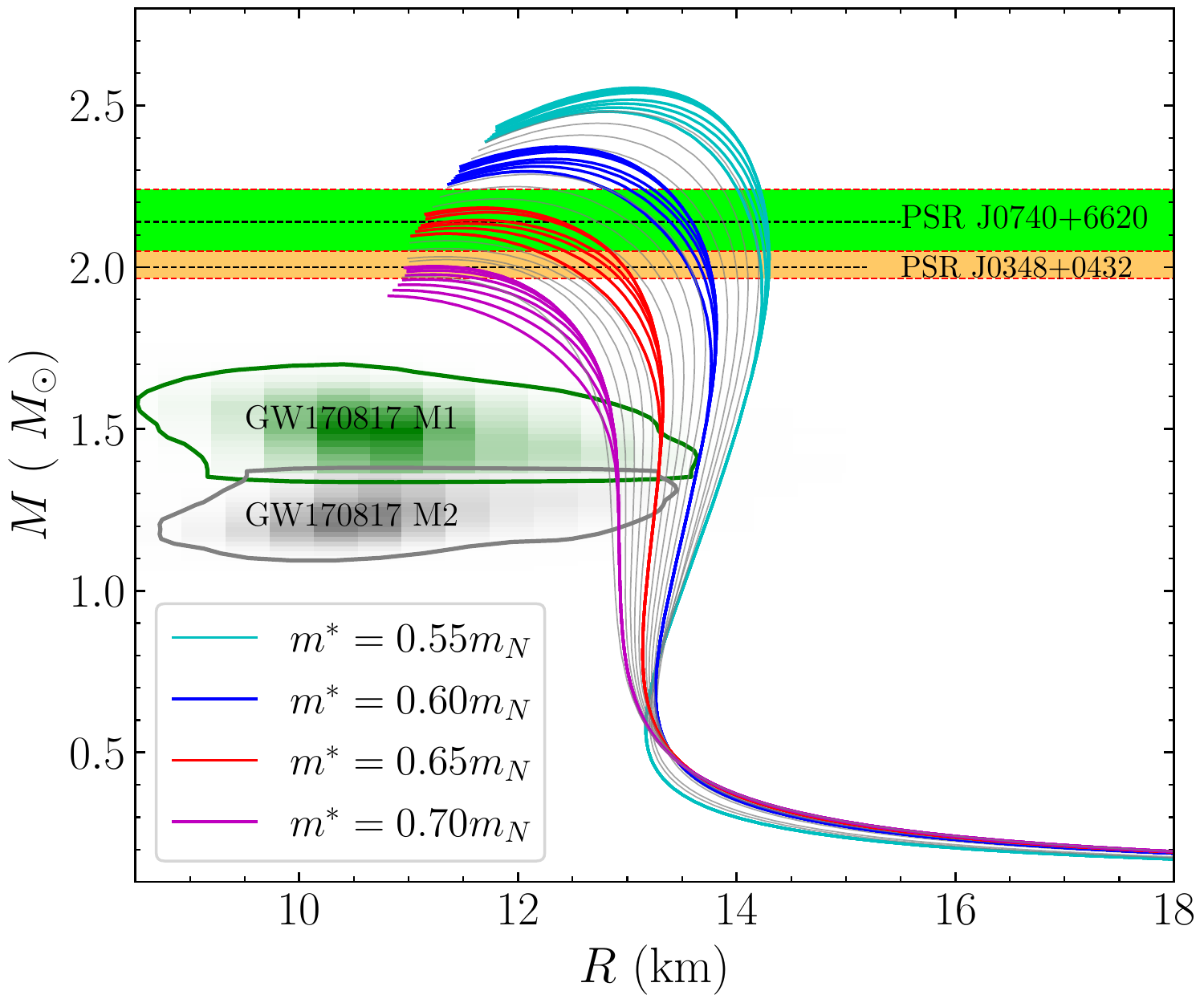}
    \caption{M-R relations corresponding to EoSs used in this work. There are two horizontal bands corresponding to pulsar masses $M=2.14^{+0.10}_{-0.09} M_{\odot}$ of PSR J0740$+$6620 \cite{Cromartie} and $M=2.01^{+0.04}_{-0.04} M_{\odot}$ of PSR J0348$+$0432 \cite{Antoniadis2013}. The mass radius estimates of the two companion neutron stars in the merger event GW170817 \cite{AbbottPRL121} are shown by shaded area labeled with GW170817 M1 (M2).
    \footnote{\url{https://dcc.ligo.org/LIGO-P1800115/public}}
    }
    \label{fig:mrgw17}
\end{figure}  

It is however interesting to note that, if one imposes additional constraints from multi-messenger astrophysical observations on the effective nucleon mass, one can restrict the physical range of $f$-mode frequencies.  We note here that recently an investigation was performed \cite{Alvarez}, where a Bayesian analysis was used to restrict the range of the Landau mass $m_L$ within an extended $\sigma-\omega$ model, using maximum NS mass measurement, tidal deformability from GW170817 and radius estimates from NICER data. As discussed earlier, if one applies the 2.1 $M_{\odot}$ maximum mass constraint, then effective nucleon mass is restricted to the range $m^*/m_N \leq 0.67$. Imposing the radius estimates from GW170817 $R_{1.4M_{\odot}} \leq$ 13.5 km (see \Cref{fig:mrgw17}) would further exclude $m^* <0.62m_N$. Similarly, one can restrict the range of $f$-mode frequencies as a function of mass from our figures. In this study we do not impose these radius constraints keeping in mind the model dependence of the results.
\\

\begin{figure}[htbp]
    \centering
    \includegraphics[width=\linewidth]{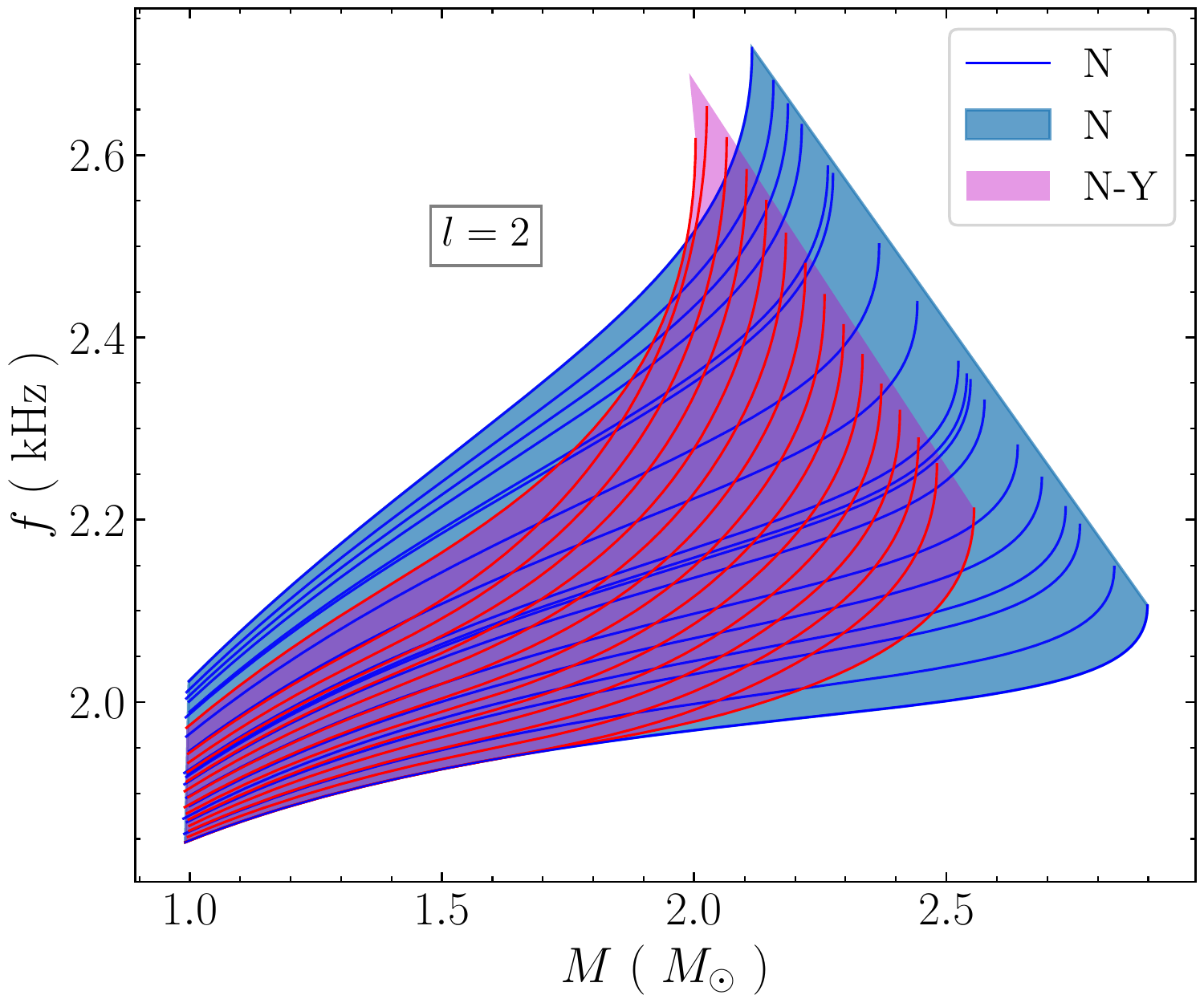}
    \caption{$f$-mode frequencies as a function of  stellar mass ($M$). The range of frequencies spanned by nucleonic cores (shaded blue) and cores with nucleons and hyperons (shaded red) are presented. Different lines with same colour represent the same system ( nucleonic core (N) or nucleon-hyperon    (N-Y)) with different parameters.}
    \label{fig:freq_massall}
\end{figure}

However it must be noted here that imposing the additional astrophysical constraints restricts the parameter space, such that the uncertainties in the nuclear saturation parameters for nucleonic and hyperonic EoSs cover similar frequency range, and do not allow us to distinguish between them on the basis of relations between $f$-mode frequencies and mass (\Cref{fig:freq_massall}), compactness (\Cref{fig:fmodecompactness}) or tidal deformability (\Cref{fig:fmodetidal}) solely. 
\\

In previous works that considered hyperons \cite{Benhar, Salcedo}, it was concluded that a relation between $f$-mode frequency and density is less useful in asteroseismology, as the calculated $f$-mode frequencies were found to be quite sensitive to the matter composition. However, in our work we performed a systematic analysis of the entire parameter range of uncertainties in the state-of-the-art nuclear and hypernuclear physics, within the framework of the RMF model. In such a case, the universal relation obtained does not vary from model to model as in previous works, as all the models chosen previously are points that lie within the considered parameter space. We found that the fit relations with and without hyperons vary slightly, given the effect of uncertainty due to hyperon potential depths is small, in comparison to the uncertainties in the nuclear saturation parameters. A more practical fit relation to apply for asteroseismology is the scaled relation between $\omega M$ and compactness $M/R$, which is independent of the EoS and quite robust. In our work, we have provided the fit relations, for $l=2$ as well as the higher order modes $l=3,4$. Using these fit relations, one may derive the mass and radius of a NS from multiple $f$-mode frequency measurements in a  model-independent way.
\\

It has been concluded that $f$-modes are among the most interesting sources of GWs due to the Chandrasekhar-Friedman-Schutz (CFS) mechanism, for both isolated NSs or in binary systems. Recent studies suggest that GWs produced by unstable $l=m=2$ and the $l=m=4$ $f$-modes could be detectable by the future Einstein Telescope for sources in the Virgo cluster or $l=m=3$ modes even by Advanced LIGO/VIRGO. It has already been estimated that  low frequency $f$-modes ($1-3 kHz$) are likely more easily observable than other modes ($p$ or $w$-modes) \cite{Lau2010}. It was discussed in \cite{Kokkotas2001} that for
a neutron star located at 10 kpc, the energy required in the $f$-mode in order to
be detected with a signal-to-noise ratio of 10 by the advanced LIGO detector would be
$8.7 \times 10^{-7} M_{\odot} $. An important breakthrough may also come with the launch of NEMO (Neutron Star Extreme Matter Observatory) \cite{NEMO}: a GW interferometer proposed by the ARC Centre of Excellence for Gravitational Wave Discovery (OzGrav) in Australia, optimised to study post-merger nuclear physics in the frequency range 2-4 kHz. \\

Several complicating effects such as rotation, magnetic fields and presence of deconfined quark core have been neglected in this work, that one has to include for a complete study of $f$-modes. Effects like these, as well as superfluidity \cite{Gualtieri2014}, would be interesting effects to investigate in future. In order to extract information from $f$-modes in the post-merger signal of a NS binary, one must include rotational effects. Superfluidity will however only play a role in cold neutron stars, as this effect will not appear in the case of $f$-modes in a hot differentially rotating remnant of a binary neutron star merger. Complementary information from GW oscillation frequencies combined with global NS properties (such as mass and tidal deformability) to high accuracy will provide an excellent tool to understand the NS core and physics of ultra-high-density matter in the unexplored regime of the QCD phase diagram. 

\section{Acknowledgements}
\label{sec:ack}
This investigation was carried out as a part of the IUCAA Graduate School project 2020. D. C. would like to thank Prashanth Jaikumar and David Alvarez-Castillo for insightful discussions. BKP is thankful to Suprovo Ghosh and Bhaskar Biswas for the useful discussion sessions they had during the project. 
\bibliography{hyp}

\end{document}